\documentclass[10pt,twocolumn,letterpaper]{article}
\pdfoutput=1
\usepackage{cvpr}
\usepackage{times}
\usepackage{epsfig}
\usepackage{graphicx}
\usepackage{amsmath}
\usepackage{amssymb}
\usepackage{subfigure}
\usepackage{cite}

\usepackage[breaklinks=true,bookmarks=false]{hyperref}

\cvprfinalcopy 

\def\cvprPaperID{6094} 

\ifcvprfinal\fi
\begin{document}

\title{M-LVC: Multiple Frames Prediction for Learned Video Compression}

\author{Jianping Lin\quad Dong Liu\thanks{This work was supported by the National Key Research and Development Program of China under Grant 2018YFA0701603, and by the Natural Science Foundation of China under Grants 61931014 and 61772483. Code and models are available at \url{https://github.com/JianpingLin/M-LVC_CVPR2020}. \emph{(Corresponding author: Dong Liu.)}}\quad Houqiang Li\quad Feng Wu\\
{\small CAS Key Laboratory of Technology in Geo-Spatial Information Processing and Application System,}\\
{\small University of Science and Technology of China, Hefei 230027, China}\\
{\tt\small ljp105@mail.ustc.edu.cn, \{dongeliu,lihq,fengwu\}@ustc.edu.cn}
}

\maketitle
\thispagestyle{empty}

\begin{abstract}
We propose an end-to-end learned video compression scheme for low-latency scenarios. Previous methods are limited in using the previous \emph{one} frame as reference. Our method introduces the usage of the previous \emph{multiple} frames as references. In our scheme, the motion vector (MV) field is calculated between the current frame and the previous one. With multiple reference frames and associated multiple MV fields, our designed network can generate more accurate prediction of the current frame, yielding less residual. Multiple reference frames also help generate MV prediction, which reduces the coding cost of MV field. We use two deep auto-encoders to compress the residual and the MV, respectively. To compensate for the compression error of the auto-encoders, we further design a MV refinement network and a residual refinement network, taking use of the multiple reference frames as well. All the modules in our scheme are jointly optimized through a single rate-distortion loss function. We use a step-by-step training strategy to optimize the entire scheme. Experimental results show that the proposed method outperforms the existing learned video compression methods for low-latency mode. Our method also performs better than H.265 in both PSNR and MS-SSIM. Our code and models are publicly available.
\end{abstract}

\section{Introduction}
Video contributes to 75\% of all Internet traffic in 2017, and the percent is expected to reach 82\% by 2022 \cite{cisco2018cisco}. Compressing video into a smaller size is an urgent requirement to reduce the transmission cost. Currently, Internet video is usually compressed into H.264 \cite{wiegand2003overview} or H.265 format \cite{sullivan2012overview}. New video coding standards like H.266 and AV1 are upcoming. While new standards promise an improvement in compression ratio, such improvement is accompanied with multiplied encoding complexity. Indeed, all the standards in use or in the way coming follow the same framework, that is motion-compensated prediction, block-based transform, and handcrafted entropy coding. The framework has been inherited for over three decades, and the development within the framework is gradually saturated.

Recently, a series of studies try to build brand-new video compression schemes on top of trained deep networks. These studies can be divided into two classes according to their targeted scenarios. As for the first class, Wu \etal proposed a recurrent neural network (RNN) based approach for interpolation-based video compression \cite{wu2018video}, where the motion information is achieved by the traditional block-based motion estimation and is compressed by an image compression method.
Later on, Djelouah \etal also proposed a method for interpolation-based video compression, where the interpolation model combines motion information compression and image synthesis, and the same auto-encoder is used for image and residual \cite{Djelouah_2019_ICCV}.
Interpolation-based compression uses the previous and the subsequent frames as references to compress the current frame, which is valid in random-access scenarios like playback.
However, it is less applicable for low-latency scenarios like live transmission.

The second class of studies target low-latency case and restrict the network to use merely temporally previous frames as references.
For example, Lu \etal proposed DVC, an end-to-end deep video compression model that jointly learns motion estimation, motion compression, motion compensation, and residual compression functions \cite{lu2018dvc}.
In this model, only one previous frame is used for motion compensation, which may not fully exploit the temporal correlation in video frames.
Rippel \etal proposed another video compression model, which maintains a latent state to memorize the information of the previous frames \cite{Rippel_2019_ICCV}. Due to the presence of the latent state, the model is difficult to train and sensitive to transmission error.

In this paper, we are interested in low-latency scenarios and propose an end-to-end learned video compression scheme. Our key idea is to use the previous \emph{multiple} frames as references. Compared to DVC, which uses only one reference frame, our used multiple reference frames enhance the prediction twofold. First, given multiple reference frames and associated multiple motion vector (MV) fields, it is possible to derive multiple hypotheses for predicting the current frame; combination of the hypotheses provides an ensemble. Second, given multiple MV fields, it is possible to extrapolate so as to predict the following MV field; using the MV prediction can reduce the coding cost of MV field. Therefore, our method is termed Multiple frames prediction for Learned Video Compression (M-LVC). Note that in \cite{Rippel_2019_ICCV}, the information of the previous multiple frames is \emph{implicitly} used to predict the current frame through the latent state; but in our scheme, the multiple frames prediction is \emph{explicitly} addressed. Accordingly, our scheme is more scalable (\ie can use more or less references), more interpretable (\ie the prediction is fulfilled by motion compensation), and easier to train per our observation.

Moreover, in our scheme, we design a MV refinement network and a residual refinement network. Since we use a deep auto-encoder to compress MV (resp. residual), the compression is lossy and incurs error in the decoded MV (resp. residual). The MV (resp. residual) refinement network is used to compensate for the compression error and to enhance the reconstruction quality. We also take use of the multiple reference frames and/or associated multiple MV fields in the residual/MV refinement network.

In summary, our technical contributions include:
\begin{itemize}
  \item
  We introduce four effective modules into end-to-end learned video compression: multiple frame-based MV prediction, multiple frame-based motion compensation, MV refinement, and residual refinement. Ablation study demonstrates the gain achieved by these modules.
  \item
  We use a single rate-distortion loss function, \emph{together with} a step-by-step training strategy, to jointly optimize all the modules in our scheme.
  \item
  We conduct extensive experiments on different datasets with various resolutions and diverse content. Our method outperforms the existing learned video compression methods for low-latency mode. Our method performs better than H.265 in both PSNR and MS-SSIM.
\end{itemize}
\vspace{-0.3cm}
\begin{figure*}
  \centering
  \subfigure[]{
    \includegraphics[width=.39\linewidth]{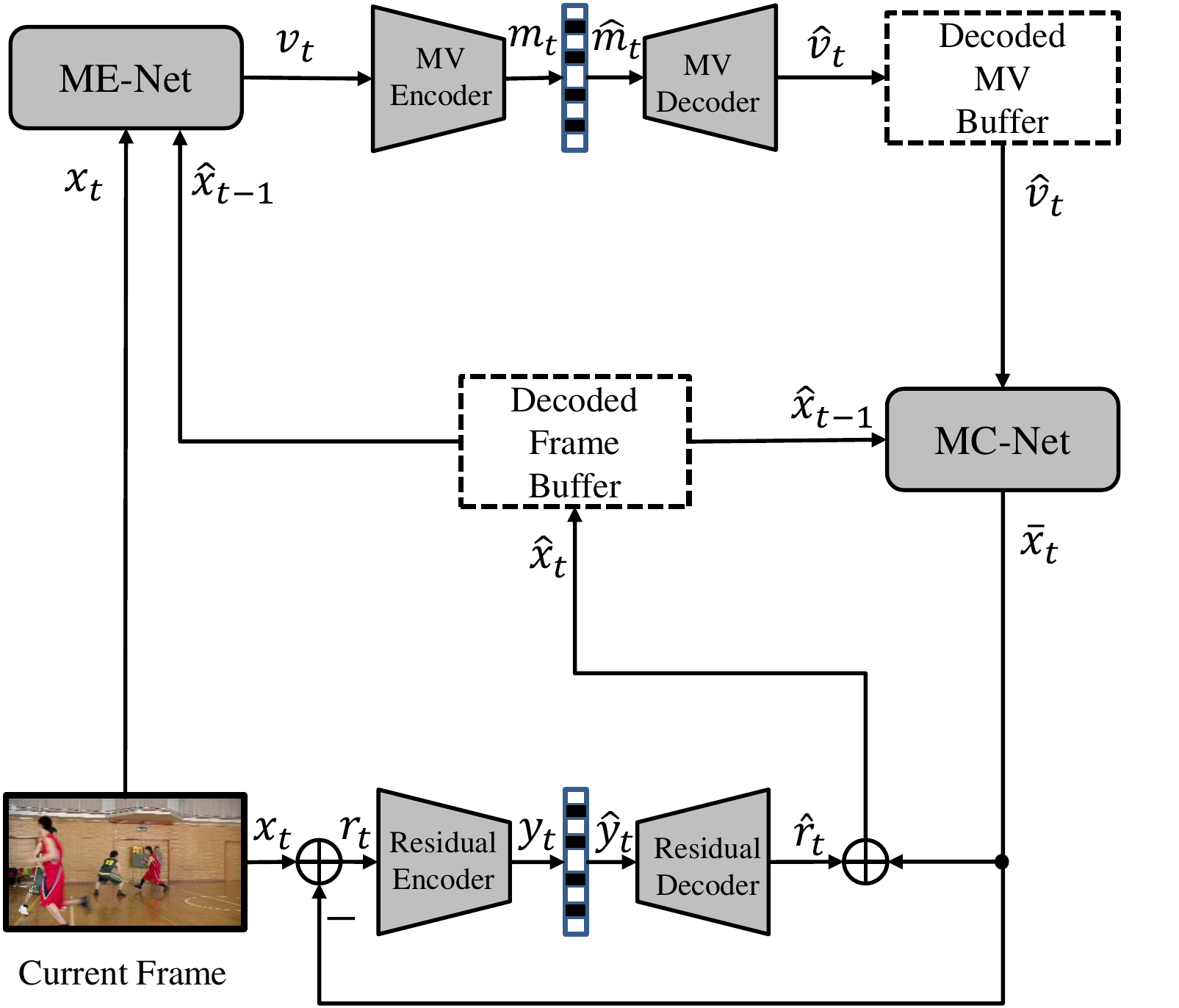}
    }
  \subfigure[]{
    \includegraphics[width=.58\linewidth]{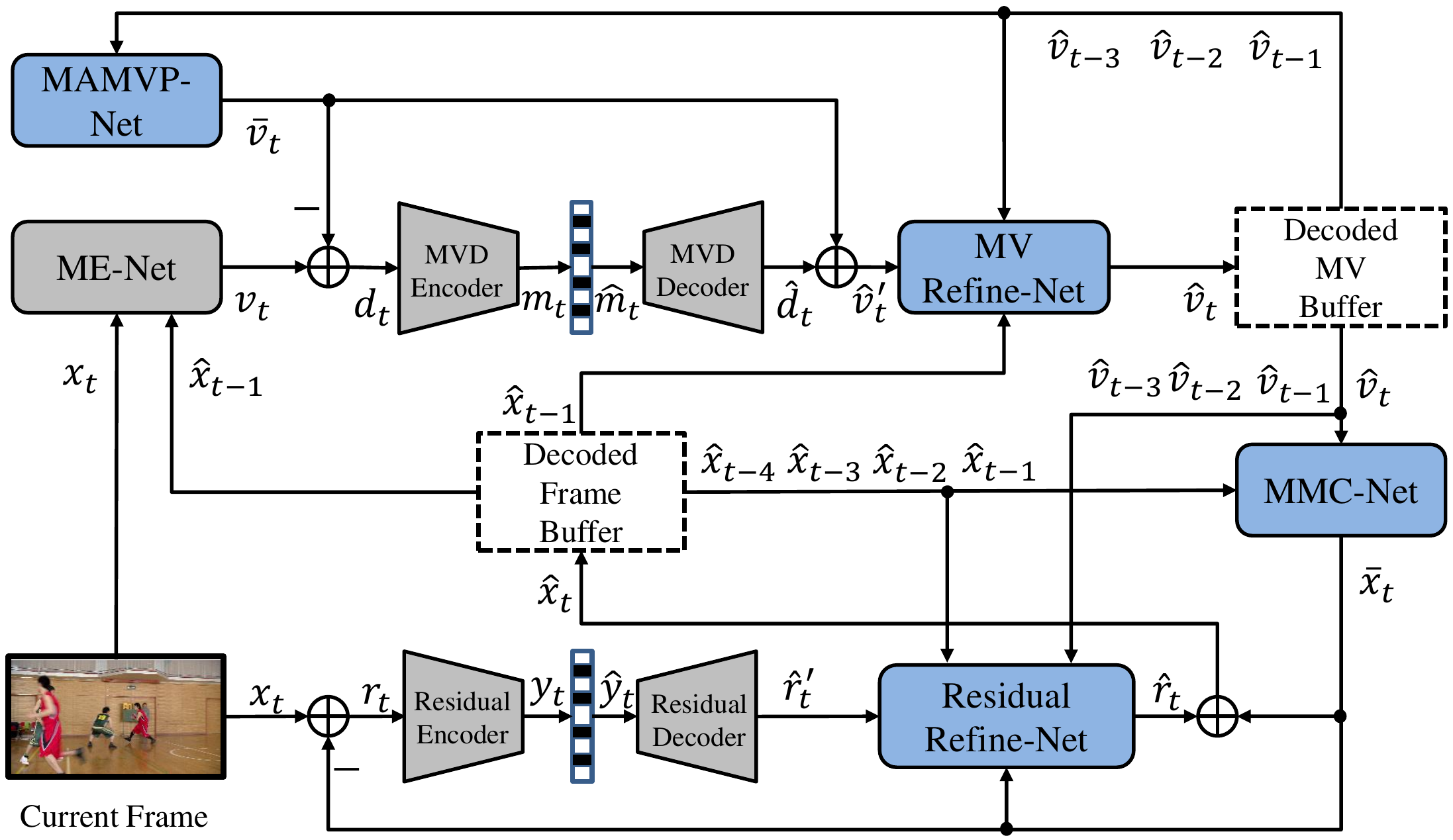}}
  \caption{(a) The scheme of DVC \cite{lu2018dvc}. (b) Our scheme. Compared to DVC, our scheme has four new modules that are highlighted in blue. In addition, our Decoded Frame Buffer stores multiple previously decoded frames as references. Our Decoded MV Buffer also stores multiple decoded MV fields. Four reference frames are depicted in the figure, which is the default setting in this paper.}
  \label{fig:framework} 
\vspace{-0.3cm}
\end{figure*}
\section{Related Work}
\subsection{Learned Image Compression}
Recently, deep learning-based image compression methods have achieved great progress \cite{johnston2018improved,toderici2015variable,toderici2017full,balle2016end,balle2018variational,minnen2018joint}. Instead of relying on handcrafted techniques like in conventional image codecs, such as JPEG \cite{wallace1992jpeg}, JPEG2000 \cite{skodras2001jpeg}, and BPG \cite{bellardbpg}, new methods can learn a non-linear transform from data and estimate the probabilities required for entropy coding in an end-to-end manner. In \cite{johnston2018improved,toderici2015variable,toderici2017full}, Long Short Term Memory (LSTM) based auto-encoders are used to progressively encode the difference between the original image and the reconstructed image. In addition, there are some studies utilizing convolutional neural network (CNN) based auto-encoders to compress images \cite{balle2016end,balle2018variational,minnen2018joint,theis2017lossy}. For example, Ball\'{e} \etal \cite{balle2016end} introduced a non-linear activation function, generalized divisive normalization (GDN), into CNN-based auto-encoder and estimated the probabilities of latent representations using a fully-connected network. This method outperformed JPEG2000. It does not take into account the input-adaptive entropy model. Ball\'{e} \etal later in \cite{balle2018variational} introduced an input-adaptive entropy model by using a zero-mean Gaussian distribution to model each latent representation and the standard deviations are predicted by a parametric transform. More recently, Minnen \etal \cite{minnen2018joint} further improved the above input-adaptive entropy model by integrating a context-adaptive model; their method outperformed BPG. In this paper, the modules for compressing the motion vector and the residual are based on the image compression methods in \cite{balle2016end,balle2018variational}. We remark that new progress on learned image compression models can be easily integrated into our scheme.
\subsection{Learned Video Compression}
Compared with learned image compression, related work for learned video compression is much less. In 2018, Wu \etal proposed a RNN-based approach for interpolation-based video compression \cite{wu2018video}. They first use an image compression model to compress the key frames, and then generate the remaining frames using hierarchical interpolation. The motion information is extracted by traditional block-based motion estimation and encoded by a traditional image compression method. Han \etal proposed to use variational auto-encoders (VAEs) for compressing sequential data \cite{han2018deep}. Their method jointly learns to transform the original video into lower-dimensional representations and to entropy code these representations according to a temporally-conditioned probabilistic model. However, their model is limited to low-resolution video. More recently, Djelouah \etal proposed a scheme for interpolation-based video compression, where the motion and blending coefficients are directly decoded from latent representations and the residual is directly computed in the latent space \cite{Djelouah_2019_ICCV}. But the interpolation model and the residual compression model are not jointly optimized.

While the above methods are designed for random-access mode, some other methods have been developed for low-latency mode. For example, Lu \etal proposed to replace the modules in the traditional video compression framework with CNN-based components, \ie motion estimation, motion compression, motion compensation, and residual compression \cite{lu2018dvc}. Their model directly compresses the motion information, and uses only one previous frame as reference for motion compensation. Rippel \etal proposed to utilize the information of multiple reference frames through maintaining a latent state \cite{Rippel_2019_ICCV}. Due to the presence of the latent state, their model is difficult to train and sensitive to transmission error. Our scheme is also tailored for low-latency mode and we will compare to \cite{lu2018dvc} more specifically in the following.

\section{Proposed Method}
{\bf Notations.}
Let $\mathcal{V}=\{x_{1},x_{2},\dots,x_{t},\dots\}$ denotes the original video sequence. $x_{t}$, $\bar{x}_{t}$, and $\hat{x}_{t}$ represent the original, predicted, and decoded/reconstructed frames at time step $t$, respectively. $r_{t}$ is the residual between the original frame $x_{t}$ and the predicted frame $\bar{x}_{t}$. $\hat{r}_{t}'$ represents the residual reconstructed by the residual auto-encoder, and $\hat{r}_{t}$ is the final decoded residual. In order to remove the temporal redundancy between video frames, we use pixel-wise motion vector (MV) field based on optical flow estimation. $v_{t}$, $\bar{v}_{t}$, and $\hat{v}_{t}$ represent the original, predicted, and decoded MV fields at time step $t$, respectively. $d_{t}$ is the MV difference (MVD) between the original MV $v_{t}$ and the predicted MV $\bar{v}_{t}$. $\hat{d}_{t}$ is the MVD reconstructed by the MVD auto-encoder, and $\hat{v}_{t}'$ represents the reconstructed MV by adding $\hat{d}_{t}$ to $\bar{v}_{t}$. Since auto-encoder represents transform, the residual $r_{t}$ and the MVD $d_{t}$ are transformed to $y_{t}$ and $m_{t}$. $\hat{y}_{t}$ and $\hat{m}_{t}$ are the corresponding quantized versions, respectively.
\subsection{Overview of the Proposed Method}
Fig.\ \ref{fig:framework} presents the scheme of DVC \cite{lu2018dvc} and our scheme for a side-by-side comparison. Our scheme introduces four new modules, which are all based on multiple reference frames. The specific compression workflow of our scheme is introduced as follows.

{\bf Step 1. Motion estimation and prediction.}
The current frame $x_{t}$ and the reference frame $\hat{x}_{t-1}$ are fed into a motion estimation network (ME-Net) to extract the motion information $v_{t}$. In this paper, the ME-Net is based on the optical flow network FlowNet2.0 \cite{ilg2017flownet}, which is at the state of the art. Instead of directly encoding the pixel-wise MV field $v_{t}$ like in Fig.\ \ref{fig:framework} (a), which incurs a high coding cost, we propose to use a MV prediction network (MAMVP-Net) to predict the current MV field, which can largely remove the temporal redundancy of MV fields. More information is provided in Section \ref{MAMVP-Net}.

{\bf Step 2. Motion compression and refinement.}
After motion prediction, we use the MVD encoder-decoder network to encode the difference $d_{t}$ between the original MV $v_{t}$ and the predicted MV $\bar{v}_{t}$.
Here the network structure is similar to that in \cite{balle2016end}.
This MVD encoder-decoder network can further remove the spatial redundancy present in $d_{t}$. Specifically, $d_{t}$ is first non-linearly mapped into the latent representations $m_{t}$, and then quantized to $\hat{m}_{t}$ by a rounding operation. The probability distributions of $\hat{m}_{t}$ are then estimated by the CNNs proposed in \cite{balle2016end}. In the inference stage, $\hat{m}_{t}$ is entropy coded into a bit stream using the estimated distributions. Then, $\hat{d}_{t}$ can be reconstructed from the entropy decoded $\hat{m}_{t}$ by the non-linear inverse transform. Since the decoded $\hat{d}_{t}$ contains error due to quantization, especially at low bit rates, we propose to use a MV refinement network (MV Refine-Net) to reduce quantization error and improve the quality. After that, the refined MV $\hat{v}_{t}$ is cached in the decoded MV buffer for next frames coding. More details are presented in Section \ref{MVR-Net}.

{\bf Step 3. Motion compensation.}
After reconstructing the MV, we use a motion compensation network (MMC-Net) to obtain the predicted frame $\bar{x}_{t}$. Instead of only using one reference frame for motion compensation like in Fig.\ \ref{fig:framework} (a), our MMC-Net can generate a more accurate prediction frame by using multiple reference frames. More information is provided in Section \ref{MC-Net}.

{\bf Step 4. Residual compression and refinement.}
After motion compensation, the residual encoder-decoder network is used to encode the residual $r_{t}$ between the original frame $x_{t}$ and the predicted frame $\bar{x}_{t}$.
The network structure is similar to that in \cite{balle2018variational}.
This residual encoder-decoder network can further remove the spatial redundancy present in $r_{t}$ by a powerful non-linear transform, which is also used in DVC \cite{lu2018dvc} because of its effectiveness. Similar to the $d_{t}$ compression, the residual $r_{t}$ is first transformed into $y_{t}$, and then quantized to $\hat{y}_{t}$. The probability distributions of $\hat{y}_{t}$ are then estimated by the CNNs proposed in \cite{balle2018variational}. In the inference stage, $\hat{y}_{t}$ is entropy coded into a bit stream using the estimated distributions. Then, $\hat{r}_{t}'$ can be reconstructed from the entropy decoded $\hat{y}_{t}$ by the non-linear inverse transform. The decoded $\hat{r}_{t}'$ contains quantization error, so we propose to use a residual refinement network (Residual Refine-Net) to reduce quantization error and enhance the quality. The details are presented in Section \ref{RR-Net}.

{\bf Step 5. Frame reconstruction.}
After refining the residual, the reconstructed frame $\hat{x}_{t}$ can be obtained by adding $\hat{r}_{t}$ to the predicted frame $\bar{x}_{t}$. $\hat{x}_{t}$ is then cached in the decoded frame buffer for next frames coding.

\subsection{Multi-scale Aligned MV Prediction Network}
\label{MAMVP-Net}
\begin{figure}
\begin{center}
\subfigure[ ]
{
\includegraphics[width=.9\linewidth]{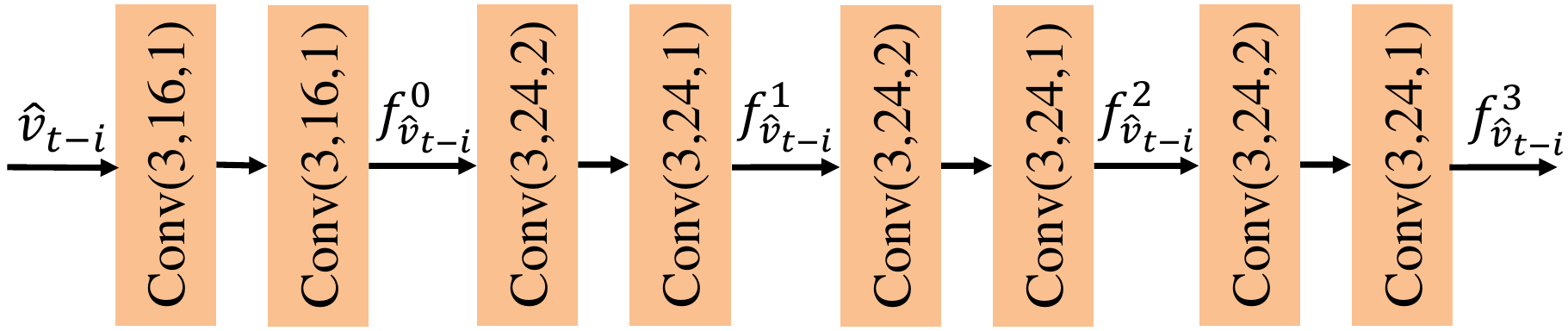}
}
\subfigure[ ]
{
\includegraphics[width=.93\linewidth]{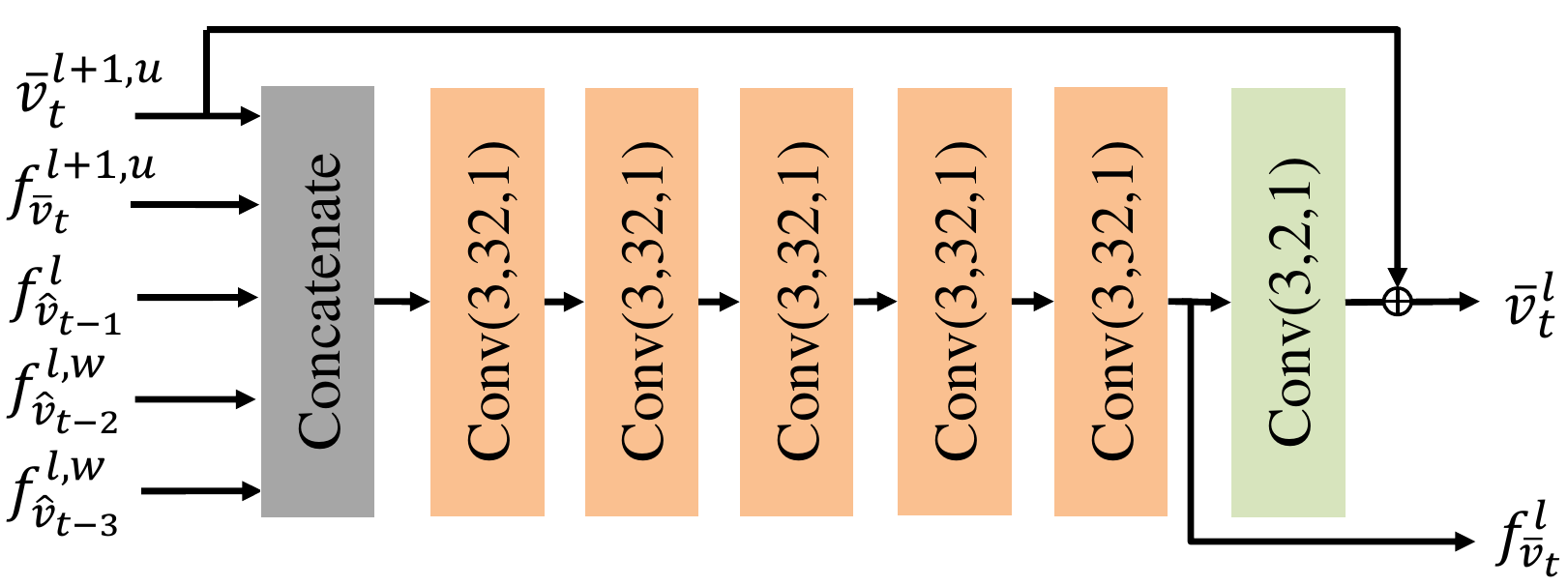}
}
\end{center}
   \caption{The multi-scale aligned MV prediction network. Conv(3,16,1) denotes the hyper-parameters of a convolutional layer: kernel size is 3$\times$3, output channel number is 16, and stride is 1. Each convolutional layer is equipped with a leaky ReLU except the one indicated by green. (a) Multi-scale feature extraction part. 2$\times$ down-sampling is performed by a convolutional layer with a stride of 2, and $i$ is 0, 1, 2. (b) MV prediction part at the $l$-th level. $l$ is 0, 1, 2, 3, and the network at the $3$-th level does not condition on the previous level.}
\label{fig:MAMVPNet}
\vspace{-0.3cm}
\end{figure}
To address large and complex motion between frames, we propose a Multi-scale Aligned MV Prediction Network (MAMVP-Net), shown in Fig.\ \ref{fig:MAMVPNet}. We use the previous three reconstructed MV fields, \ie $\hat{v}_{t-3}$, $\hat{v}_{t-2}$, and $\hat{v}_{t-1}$, to obtain the MV prediction $\bar{v}_{t}$. More or less MV fields may be used depending on the size of the Decoded MV Buffer.

As shown in Fig.\ \ref{fig:MAMVPNet} (a), we first generate a multi-level feature pyramid for each previous reconstructed MV field, using a multi-scale feature extraction network (four levels are used for example),
\vspace{-0.07cm}
 \begin{equation}\label{multi-scale_extractor}
  \{f_{\hat{v}_{t-i}}^{l}|l=0,1,2,3\}=H_{mf}(\hat{v}_{t-i}), i=1,2,3
\end{equation}
where $f_{\hat{v}_{t-i}}^{l}$ represents the features of $\hat{v}_{t-i}$ at the $l$-th level.
Second, considering the previous reconstructed MV fields contain compression error, we choose to warp the feature pyramids of $\hat{v}_{t-3}$ and $\hat{v}_{t-2}$, instead of the MV fields themselves, towards $\hat{v}_{t-1}$ via:
\vspace{-0.1cm}
\begin{equation}\label{warp_mvp}
\begin{split}
f_{\hat{v}_{t-3}}^{l,w} &= Warp(f_{\hat{v}_{t-3}}^{l},\hat{v}_{t-1}^{l}+Warp(\hat{v}_{t-2}^{l},\hat{v}_{t-1}^{l})) \\
f_{\hat{v}_{t-2}}^{l,w} &= Warp(f_{\hat{v}_{t-2}}^{l},\hat{v}_{t-1}^{l}),l=0,1,2,3
\end{split}
\end{equation}
where $f_{\hat{v}_{t-3}}^{l,w}$ and $f_{\hat{v}_{t-2}}^{l,w}$ are the warped features of $\hat{v}_{t-3}$ and $\hat{v}_{t-2}$ at the $l$-th level. $\hat{v}_{t-1}^{l}$ and $\hat{v}_{t-2}^{l}$ are the down-sampled versions of $\hat{v}_{t-1}$ and $\hat{v}_{t-2}$ at the $l$-th level. $Warp$ stands for bilinear interpolation-based warping. Note that feature domain warping has been adopted in previous work because of its effectiveness, such as in \cite{niklaus2018context} for video frame interpolation and in \cite{sun2018pwc} for optical flow generation.
Third, we use a pyramid network to predict the current MV field from coarse to fine based on the feature pyramid of $\hat{v}_{t-1}$ and the warped feature pyramids of $\hat{v}_{t-2}$ and $\hat{v}_{t-3}$. As shown in Fig.\ \ref{fig:MAMVPNet} (b), the predicted MV field $\bar{v}^{l}_{t}$ and the predicted features $f_{\bar{v}_{t}}^{l}$ at the $l$-th level can be obtained via:
\begin{equation}\label{predict_mvp}
  \bar{v}^{l}_{t}, f_{\bar{v}_{t}}^{l} = H_{mvp}(\bar{v}^{l+1,u}_{t},f_{\bar{v}_{t}}^{l+1,u},f_{\hat{v}_{t-1}}^{l},f_{\hat{v}_{t-2}}^{l,w},f_{\hat{v}_{t-3}}^{l,w})
\end{equation}
where $\bar{v}^{l+1,u}_{t}$ and $f_{\bar{v}_{t}}^{l+1,u}$ are the 2$\times$ up-sampled MV field and features from those at the previous $(l+1)$-th level using bilinear. This process is repeated until the desired $0$-th level, resulting in the final predicted MV field $\bar{v}_{t}$.

\subsection{MV Refinement Network}
\label{MVR-Net}
After MVD compression, we can reconstruct the MV field $\hat{v}_{t}'$ by adding the decoded MVD $\hat{d}_{t}$ to the predicted MV $\bar{v}_{t}$. But $\hat{v}_{t}'$ contains compression error caused by quantization, especially at low bit rates. For example, we found there are many zeros in $\hat{d}_{t}$, as zero MVD requires less bits to encode. A similar result is also reported in DVC \cite{lu2018dvc} when compressing the MV field. But such zero MVD incurs inaccurate motion compensation. Therefore, we propose to use a MV refinement network (MV Refine-Net) to reduce compression error and improve the accuracy of reconstructed MV. As shown in Fig.\ \ref{fig:framework} (b), we use the previous three reconstructed MV fields, \ie $\hat{v}_{t-3}$, $\hat{v}_{t-2}$, and $\hat{v}_{t-1}$, and the reference frame $\hat{x}_{t-1}$ to refine $\hat{v}_{t}'$. Using the previous multiple reconstructed MV fields can more accurately predict the current MV, and then help on refinement. The reason for using $\hat{x}_{t-1}$ is that the following motion compensation module will depend on the refined $\hat{v}_{t}$ and $\hat{x}_{t-1}$ to obtain the predicted frame, so $\hat{x}_{t-1}$ can be a guidance to help refine $\hat{v}_{t}'$. According to our experimental results (Section \ref{Ablation}), feeding $\hat{x}_{t-1}$ into the MV refinement network does improve the compression efficiency. More details of the MV Refine-Net can been found in the supplementary.

\subsection{Motion Compensation Network with Multiple Reference Frames}
\label{MC-Net}
\begin{figure}
\begin{center}
\includegraphics[width=.85\linewidth]{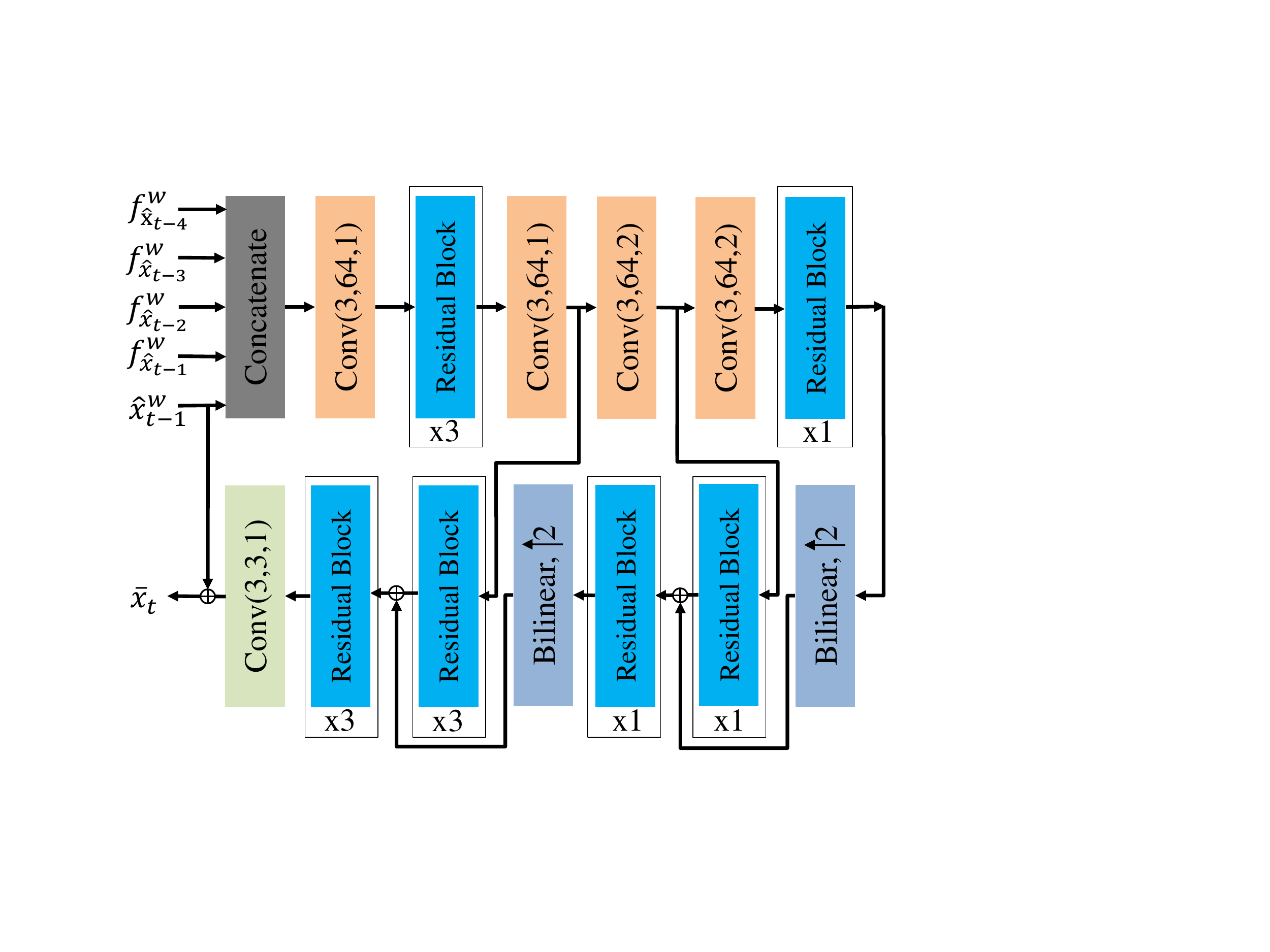}
\end{center}
   \caption{The motion compensation network. Each convolutional layer outside residual blocks is equipped with a leaky ReLU except the last layer (indicated by green). Each residual block consists of two convolutional layers, which are configured as follows: kernel size is 3$\times$3, output channel number is 64, the first layer has ReLU.}
\label{fig:MCNet}
\vspace{-0.3cm}
\end{figure}
In traditional video coding schemes, the motion compensation using multiple reference frames is adopted in H.264/AVC \cite{wiegand2003overview}, and inherited by the following standards. For example, some coding blocks may use a weighted average of two different motion-compensated predictions from different reference frames, which greatly improves the compression efficiency. Besides, in recent work for video super-resolution, multiple frames methods are also observed much better than those based on a single frame \cite{wang2019edvr,Haris_2019_CVPR,Li_2019_CVPR}. Therefore, we propose to use multiple reference frames for motion compensation in our scheme.

The network architecture is shown in Fig.\ \ref{fig:MCNet}. In this module, we use the previous four reference frames, \ie $\hat{x}_{t-4}$, $\hat{x}_{t-3}$, $\hat{x}_{t-2}$ and $\hat{x}_{t-1}$ to obtain the predicted frame $\bar{x}_{t}$. More or less reference frames can be used depending on the size of the Decoded Frame Buffer.
First, we use a two-layer CNN to extract the features of each reference frame. Then, the extracted features and $\hat{x}_{t-1}$ are warped towards the current frame via:
\vspace{-0.3cm}
\begin{equation}\label{warp_frames}
\begin{split}
  \hat{v}^{w}_{t-k} &= Warp(\hat{v}_{t-k},\hat{v}_{t}+\sum_{l=1}^{k-1}\hat{v}^{w}_{t-l}), k=1,2,3\\
  \hat{x}^{w}_{t-1} &= Warp(\hat{x}_{t-1},\hat{v}_{t})\\
  f_{\hat{x}_{t-i}}^{w} &= Warp(f_{\hat{x}_{t-i}},\hat{v}_{t}+\sum_{k=1}^{i-1}\hat{v}^{w}_{t-k}), i=1,2,3,4 \\
\end{split}
\end{equation}
where $\hat{v}^{w}_{t-k}$ is the warped version of $\hat{v}_{t-k}$ towards $\hat{v}_{t}$, and $f_{\hat{x}_{t-i}}^{w}$ is the warped feature of $\hat{x}_{t-i}$. Finally, as Fig.\ \ref{fig:MCNet} shows, the warped features and frames are fed into a CNN to obtain the predicted frame,
\begin{equation}\label{res-equation}
  \bar{x}_{t} = H_{mc}(f_{\hat{x}_{t-4}}^{w}, f_{\hat{x}_{t-3}}^{w}, f_{\hat{x}_{t-2}}^{w}, f_{\hat{x}_{t-1}}^{w}, \hat{x}^{w}_{t-1}) + \hat{x}^{w}_{t-1}
\end{equation}
where the network is based on the U-Net structure \cite{ronneberger2015u} and integrates multiple residual blocks.

\subsection{Residual Refinement Network}
\label{RR-Net}
After residual compression, the reconstructed residual $\hat{r}_{t}'$ contains compression error, especially at low bit rates. Similar to the case of MV Refine-Net, we propose a residual refinement network (Residual Refine-Net) to reduce compression error and improve quality. As shown in Fig.\ \ref{fig:framework} (b), this module utilizes the previous four reference frames, \ie $\hat{x}_{t-4}$, $\hat{x}_{t-3}$, $\hat{x}_{t-2}$ and $\hat{x}_{t-1}$, and the predicted frame $\bar{x}_{t}$ to refine $\hat{r}_{t}'$. More details of this network are provided in the supplementary.
\subsection{Training Strategy}
\label{Training_strategy}
{\bf Loss Function.}
Our scheme aims to jointly optimize the number of encoding bits and the distortion between the original frame $x_{t}$ and the reconstructed frame $\hat{x}_{t}$. We use the following loss function for training,
\begin{equation}\label{loss-equation}
  J = D + \lambda R = d(x_{t},\hat{x}_{t}) + \lambda (R_{mvd}+R_{res})
\end{equation}
where $d(x_{t},\hat{x}_{t})$ is the distortion between $x_{t}$ and $\hat{x}_{t}$. We use the mean squared error (MSE) as distortion measure in our experiments. $R_{mvd}$ and $R_{res}$ represent the bit rates used for encoding the MVD $d_{t}$ and the residual $r_{t}$, respectively. During training, we do not perform real encoding but instead estimate the bit rates from the entropy of the corresponding latent representations $\hat{m}_{t}$ and $\hat{y}_{t}$. We use the CNNs in \cite{balle2016end} and \cite{balle2018variational} to estimate the probability distributions of $\hat{m}_{t}$ and $\hat{y}_{t}$, respectively, and then obtain the corresponding entropy. Since $\hat{m}_{t}$ and $\hat{y}_{t}$ are the quantized representations and the quantization operation is not differentiable, we use the method proposed in \cite{balle2016end}, where the quantization operation is replaced by adding uniform noise during training.
\begin{figure*}
  \centering
  \subfigure[]{
    \includegraphics[width=2.0in]{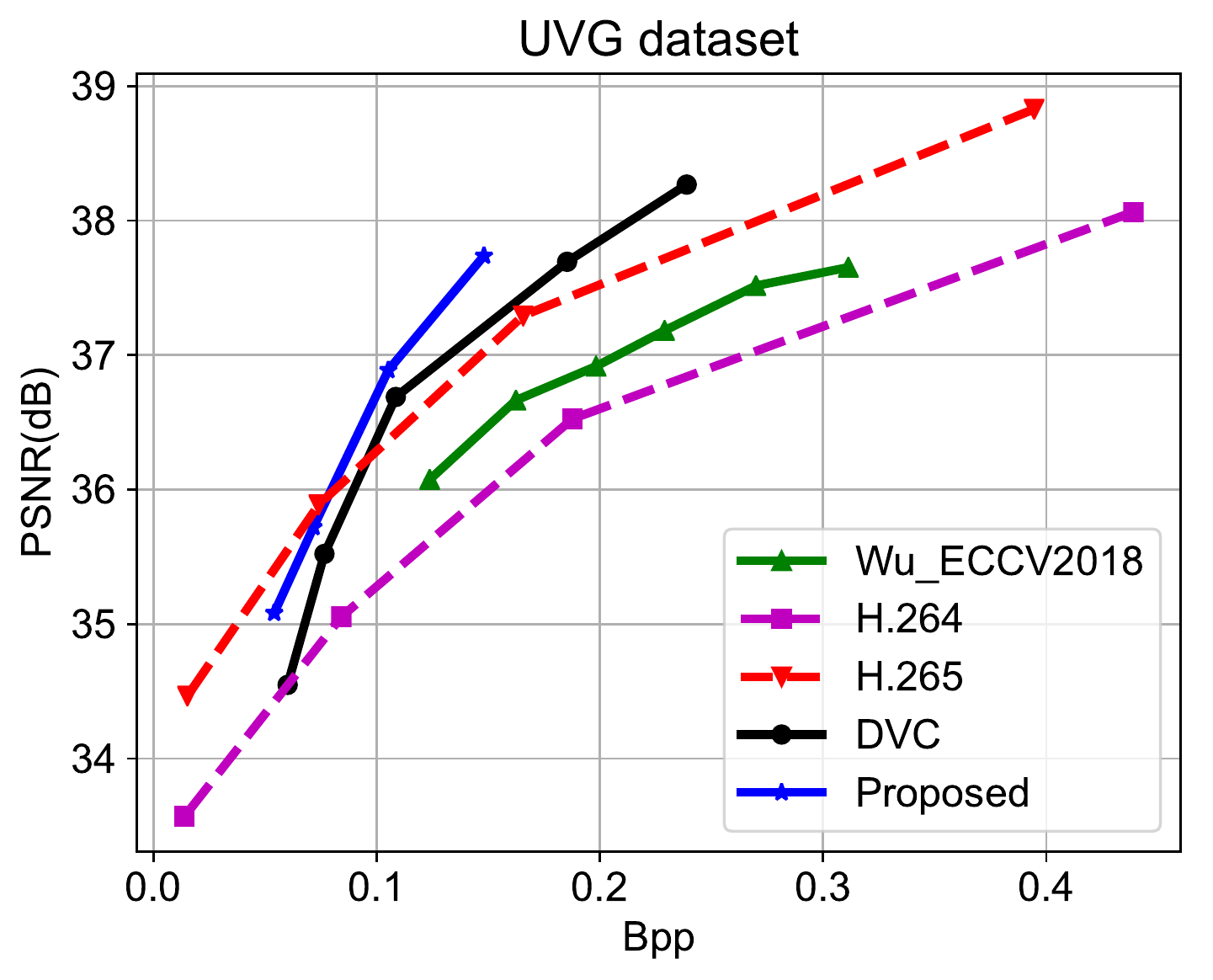}}
  \subfigure[]{
    \includegraphics[width=2.0in]{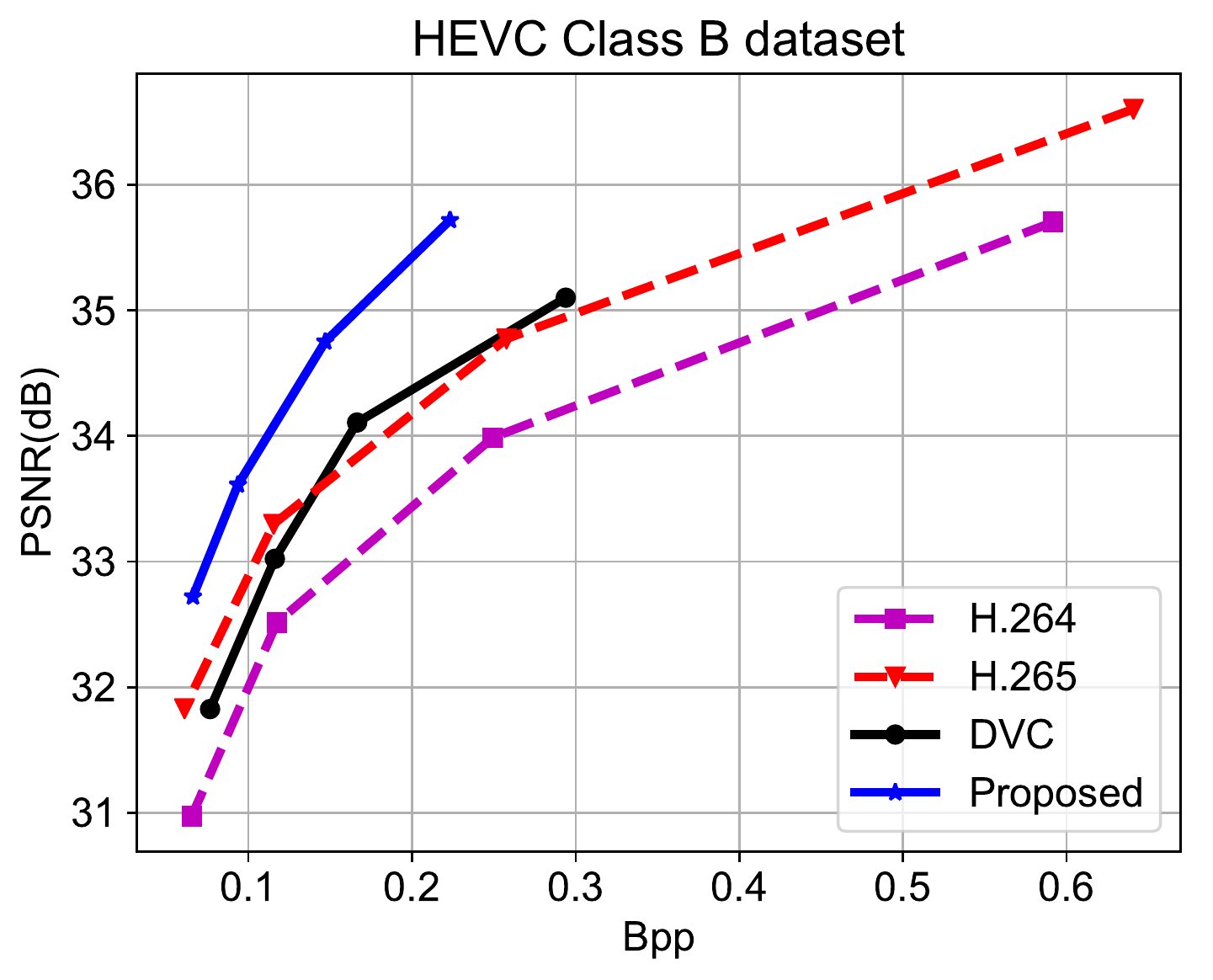}}
  \subfigure[]{
    \includegraphics[width=2.0in]{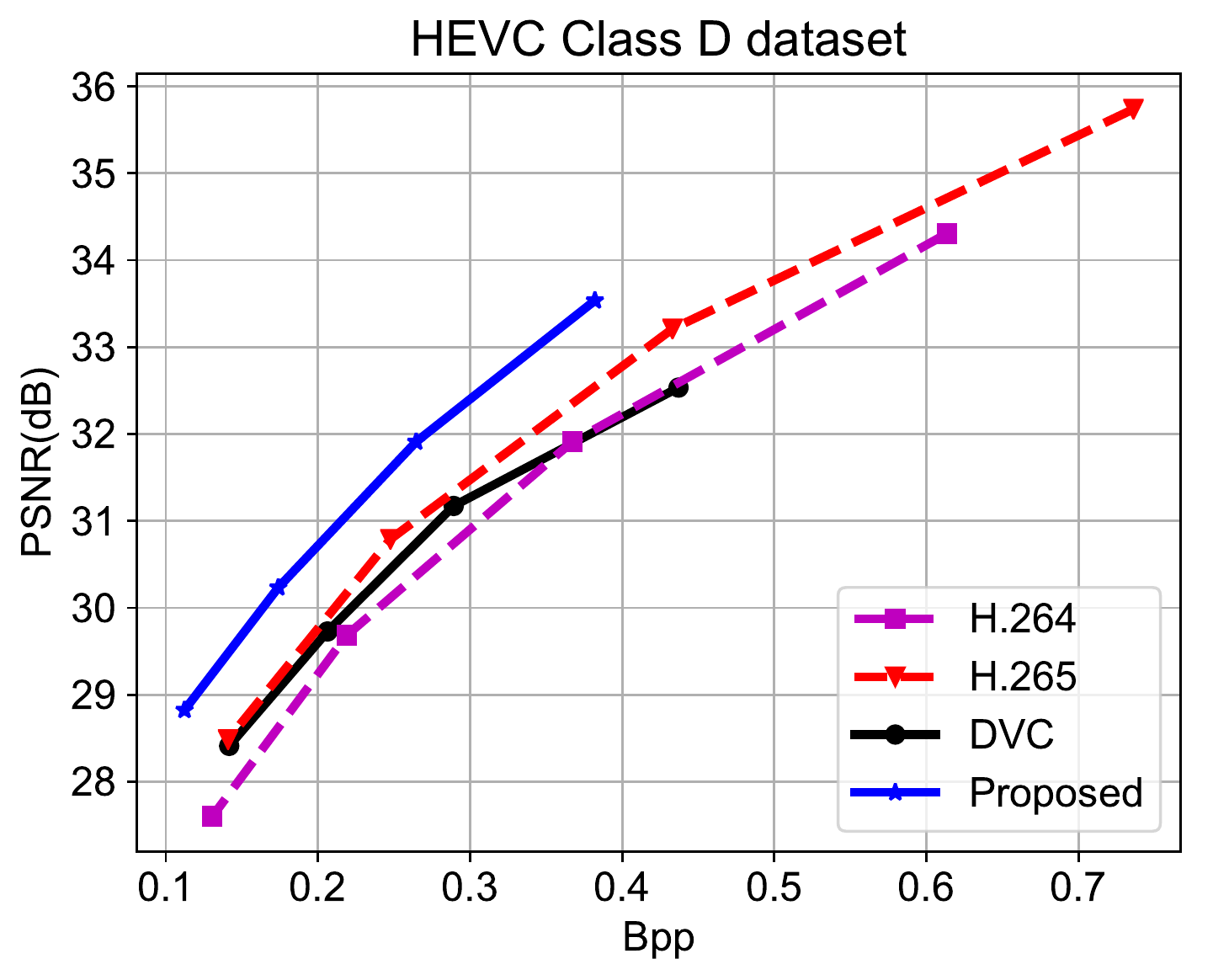}}
  \subfigure[]{
    \includegraphics[width=2.0in]{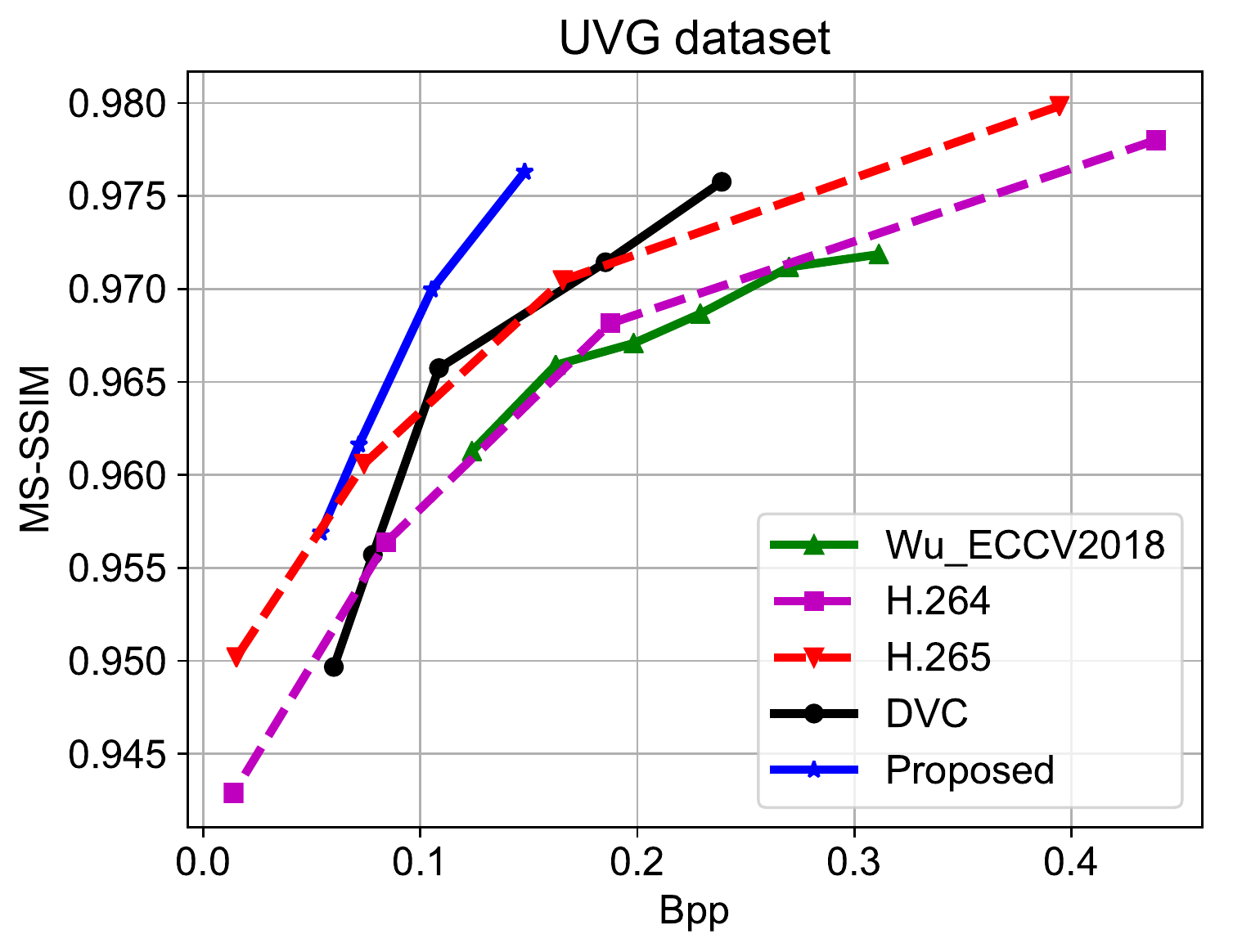}}
  \subfigure[]{
    \includegraphics[width=2.0in]{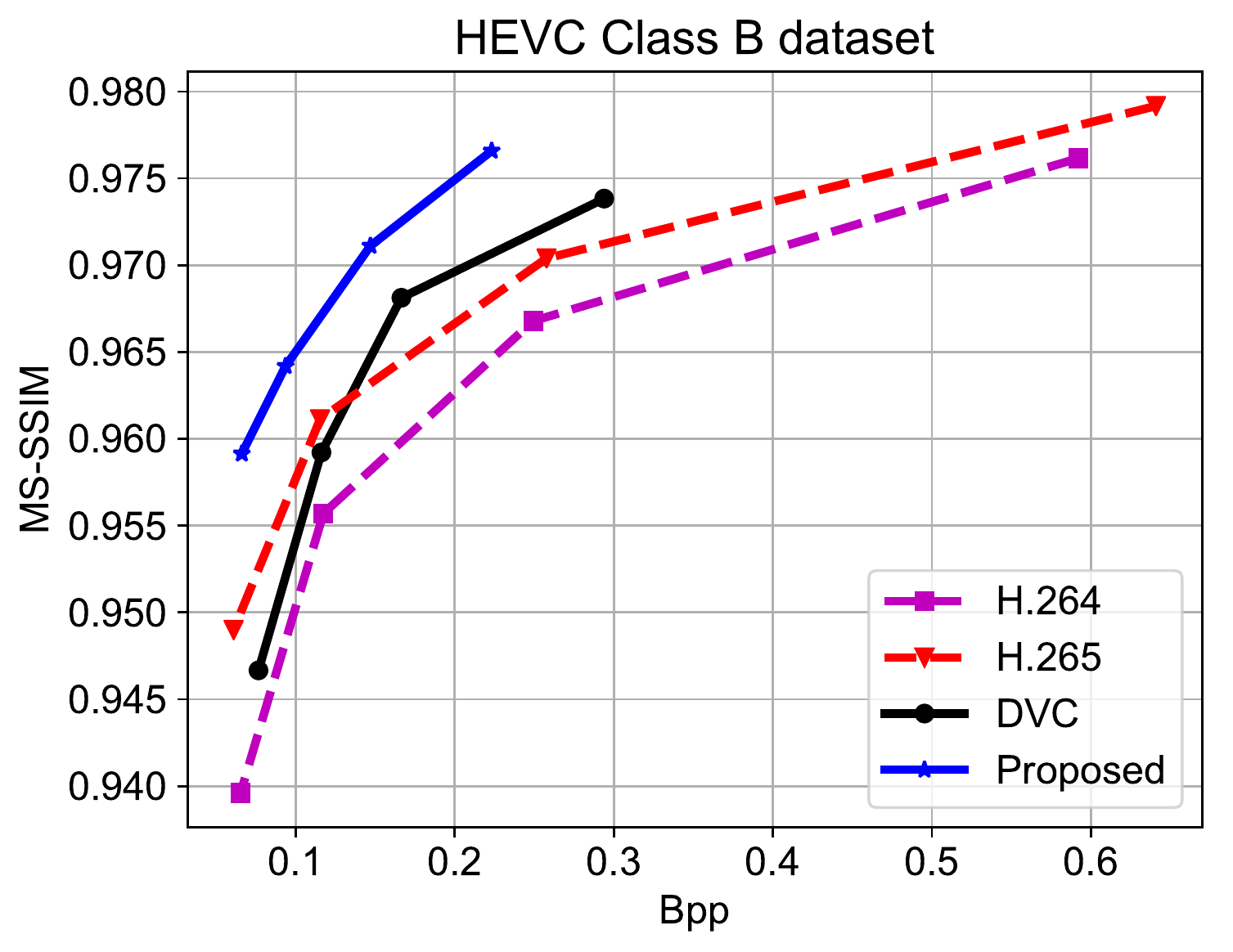}}
  \subfigure[]{
    \includegraphics[width=2.0in]{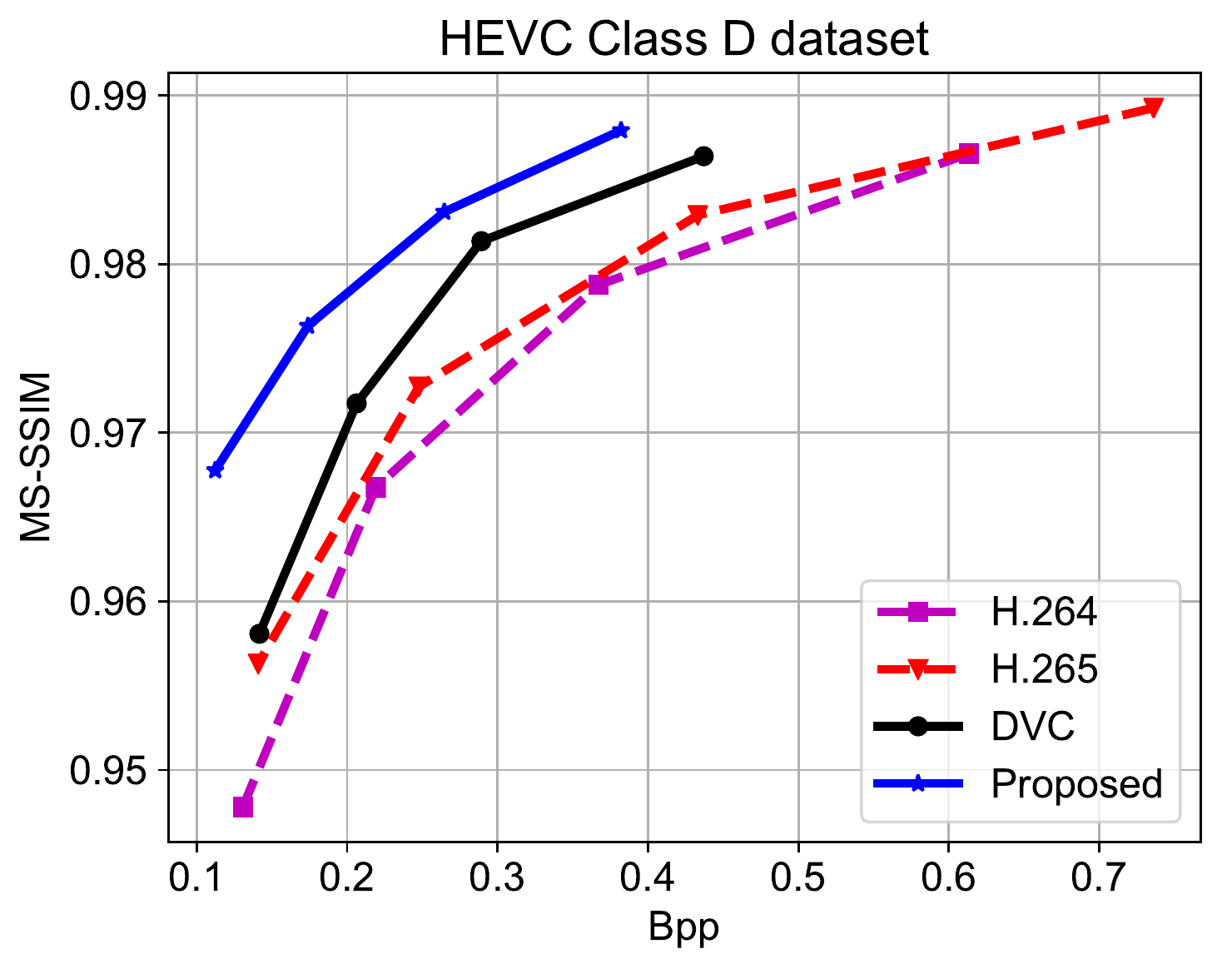}}
  \caption{\textbf{Overall performance}. The compression results on the three datasets using H.264 \cite{wiegand2003overview}, H.265 \cite{sullivan2012overview}, DVC \cite{lu2018dvc}, Wu's method \cite{wu2018video} and the proposed method. We directly use the results reported in \cite{lu2018dvc} and \cite{wu2018video}. The results of H.264 and H.265 are cited from \cite{lu2018dvc}. Wu \cite{wu2018video} did not report on HEVC Class B and Class D. Top row: PSNR. Bottom row: MS-SSIM.}
  \label{fig_RD Curve} 
\vspace{-0.3cm}
\end{figure*}

{\bf Progressive Training.}
We had tried to train the entire network from scratch, \ie with all the modules except the ME-Net randomly initialized (ME-Net is readily initialized with FlowNet2.0). The results are not satisfactory, as the resulting bitrates are not balanced: too less rate for MVD and too much rate for residual, and thus the compression results are inefficient (see the experimental results in Section \ref{Ablation}).
To address this problem, we use a step-by-step training strategy. First, we train the network including only the ME-Net and MMC-Net, while the ME-Net is the pre-trained model in \cite{ilg2017flownet} and remains unchanged. Then, the MVD and residual encoder-decoder networks are added for training, while the parameters of ME-Net and MMC-Net are fixed. After that, all of the above four modules are jointly fine-tuned. Next, we add the MAMVP-Net, MV Refine-Net and Residual Refine-Net one by one to the training system. Each time when adding a new module, we fix the previously trained modules and learn the new module specifically, and then jointly fine-tune all of them. It is worth noting that many previous studies that use step-by-step training usually adopt a different loss function for each step (\eg \cite{reda2018sdc,Yang_2018_CVPR}), while the loss function remains the same rate-distortion cost in our method.
\vspace{-0.08cm}
\section{Experiments}
\subsection{Experimental Setup}
\vspace{-0.08cm}
{\bf Training Data.}
We use the Vimeo-90k dataset \cite{xue2019video}, and crop the large and long video sequences into 192$\times$192, 16-frame video clips.

{\bf Implementation Details.}
In our experiments, the coding structure is IPPP\dots~and all the P-frames are compressed by the same network. We do not implement a single image compression network but use H.265 to compress the only I-frame. For the first three P-frames, whose reference frames are less than four, we duplicate the furthest reference frame to achieve the required four frames. We train four models with different $\lambda$ values ($16, 24, 40, 64$) for multiple coding rates. The Adam optimizer \cite{kingma2014adam} with the momentum of $0.9$ is used. The initial learning rate is $5e{-5}$ for training newly added modules, and $1e{-5}$ in the fine-tuning stages. The learning rate is reduced by a factor of $2$ five times during training. Batch size is $8$ (\ie $8$ cropped clips). The entire scheme is implemented by TensorFlow and trained/tested on a single Titan Xp GPU.

{\bf Testing Sequences.}
The HEVC common test sequences, including 16 videos of different resolutions known as Classes B, C, D, E \cite{bossen2011common}, are used for evaluation. We also use the seven sequences at 1080p from the UVG dataset \cite{uvgdata}.

{\bf Evaluation Metrics.}
Both PSNR and MS-SSIM \cite{wang2003multiscale} are used to measure the quality of the reconstructed frames in comparison to the original frames. Bits per pixel (bpp) is used to measure the number of bits for encoding the representations including MVD and residual.
\vspace{-0.08cm}
\subsection{Experimental Results}
To demonstrate the advantage of our proposed scheme, we compare with existing video codecs, in particular H.264 \cite{wiegand2003overview} and H.265 \cite{sullivan2012overview}. For easy comparison with DVC, we directly cite the compression results of H.264 and H.265 reported in \cite{lu2018dvc}. The results of H.264 and H.265 default settings can be found in the supplementary.

In addition, we compare with several state-of-the-art learned video compression methods, including Wu\_ECCV2018 \cite{wu2018video} and DVC \cite{lu2018dvc}. To the best of our knowledge, DVC \cite{lu2018dvc} reports the best compression performance in PSNR among the learning-based methods for low-latency mode.

Fig.\ \ref{fig_RD Curve} presents the compression results on the UVG dataset and the HEVC Class B and Class D datasets. It can be observed that our method outperforms the learned video compression methods DVC \cite{lu2018dvc} and Wu\_ECCV2018 \cite{wu2018video} by a large margin. On the HEVC Class B dataset, our method achieves about 1.2dB coding gain than DVC at the same bpp of 0.226. When compared with the traditional codec H.265, our method has achieved better compression performance in both PSNR and MS-SSIM. The gain in MS-SSIM seems more significant. It is worth noting that our model is trained with the MSE loss, but results show that it also works for MS-SSIM.
More experimental results, including HEVC Class C and Class E, comparisons to other methods \cite{Djelouah_2019_ICCV,Rippel_2019_ICCV}, are given in the supplementary.\vspace{-0.08cm}

\subsection{Ablation Study}
\label{Ablation}
\begin{figure}
\begin{center}
\includegraphics[width=.8\linewidth]{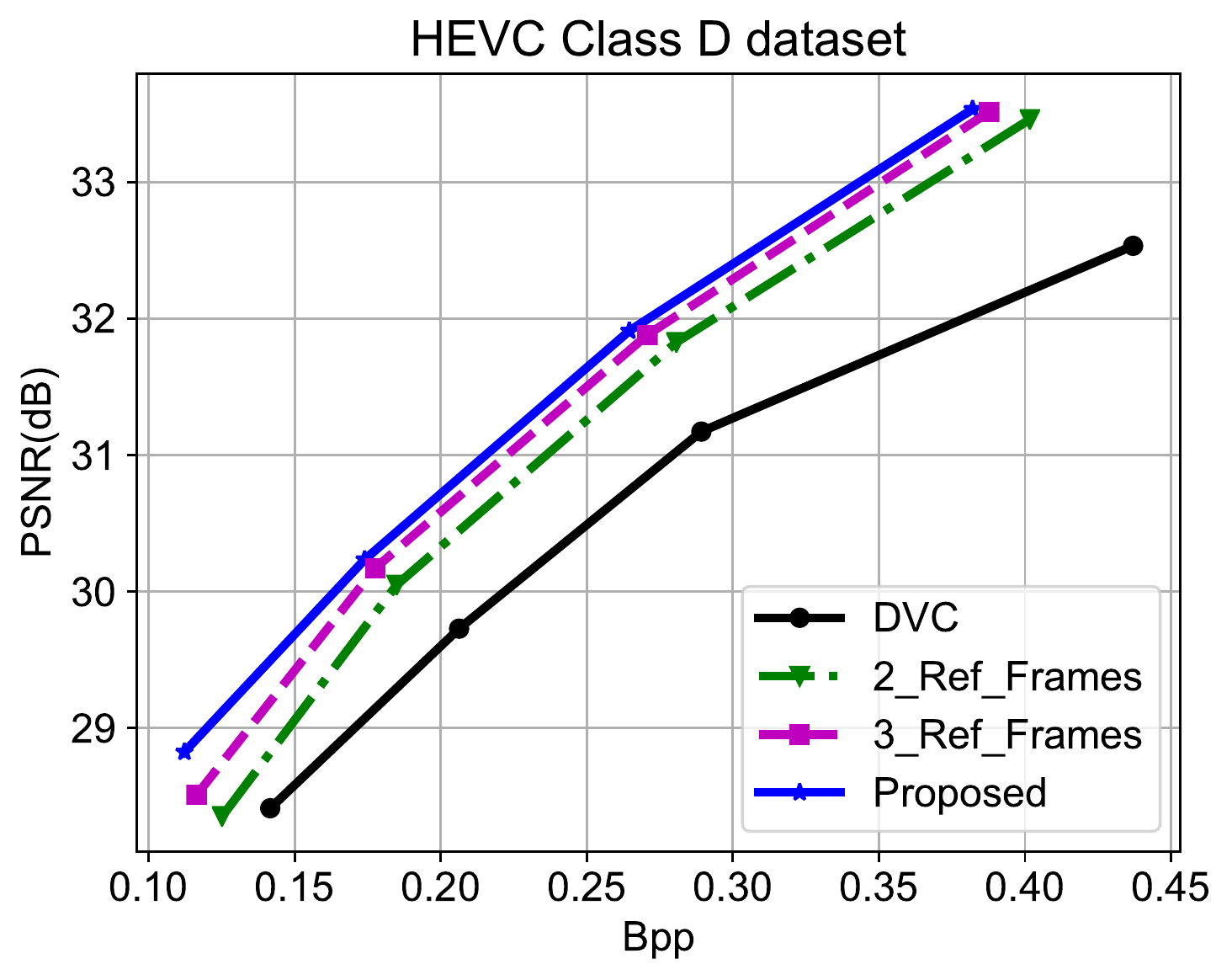}
\end{center}
   \caption{The compression results of using two or three reference frames in our trained models on the HEVC Class D dataset. The proposed model uses four by default and DVC \cite{lu2018dvc} uses only one.}
\label{fig:RF_Ablation}
\vspace{-0.5cm}
\end{figure}

\begin{figure}
\begin{center}
\includegraphics[width=.8\linewidth]{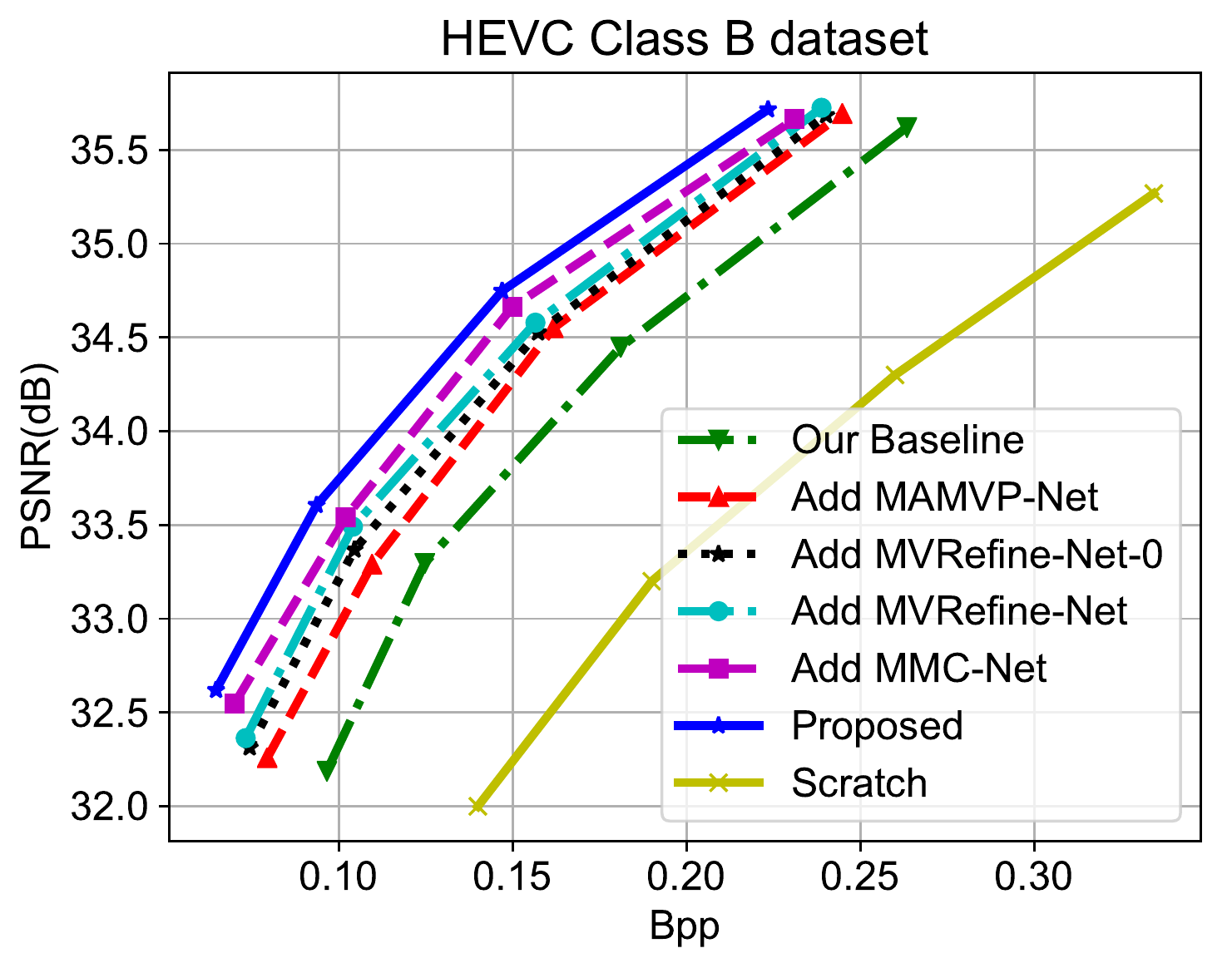}
\end{center}
   \caption{{\bf Ablation study.} The compression results of the following settings on the HEVC Class B dataset. (1) \texttt{Our Baseline}: The network contains ME-Net, MC-Net with only one reference frame, and the MV and residual encoder-decoder networks. (2) \texttt{Add MAMVP-Net}: The MAMVP-Net is added to (1). (3) \texttt{Add MVRefine-Net}: The MV Refine-Net is added to (2). (4) \texttt{Add MVRefine-Net-0}: $f_{\hat{x}_{t-1}}$ is removed from the MV Refine-Net in (3). (5) \texttt{Add MMC-Net}: The MC-Net with one reference frame in (3) is replaced by the MMC-Net with multiple reference frames. (6) \texttt{Proposed}: The Residual Refine-Net is added to (5). (7) \texttt{Scratch}: Training (6) from scratch.}
\label{fig:Ablation}
\vspace{-0.55cm}
\end{figure}
{\bf On the Number of Reference Frames.}
The number of reference frames is an important hyper-parameter in our scheme. Our used default value is four reference frames and their associated MV fields, which is also the default value in the H.265 reference software. To evaluate the effectiveness of using less reference frames, we conduct a comparison experiment by using two or three reference frames in our trained models. Fig.\ \ref{fig:RF_Ablation} presents the compression results on the HEVC Class D dataset. As observed, the marginal gain of increasing reference frame is less and less.

{\bf Multi-scale Aligned MV Prediction Network.}
\begin{figure*}
  \centering
  \subfigure[]{
    \includegraphics[width=.17\linewidth]{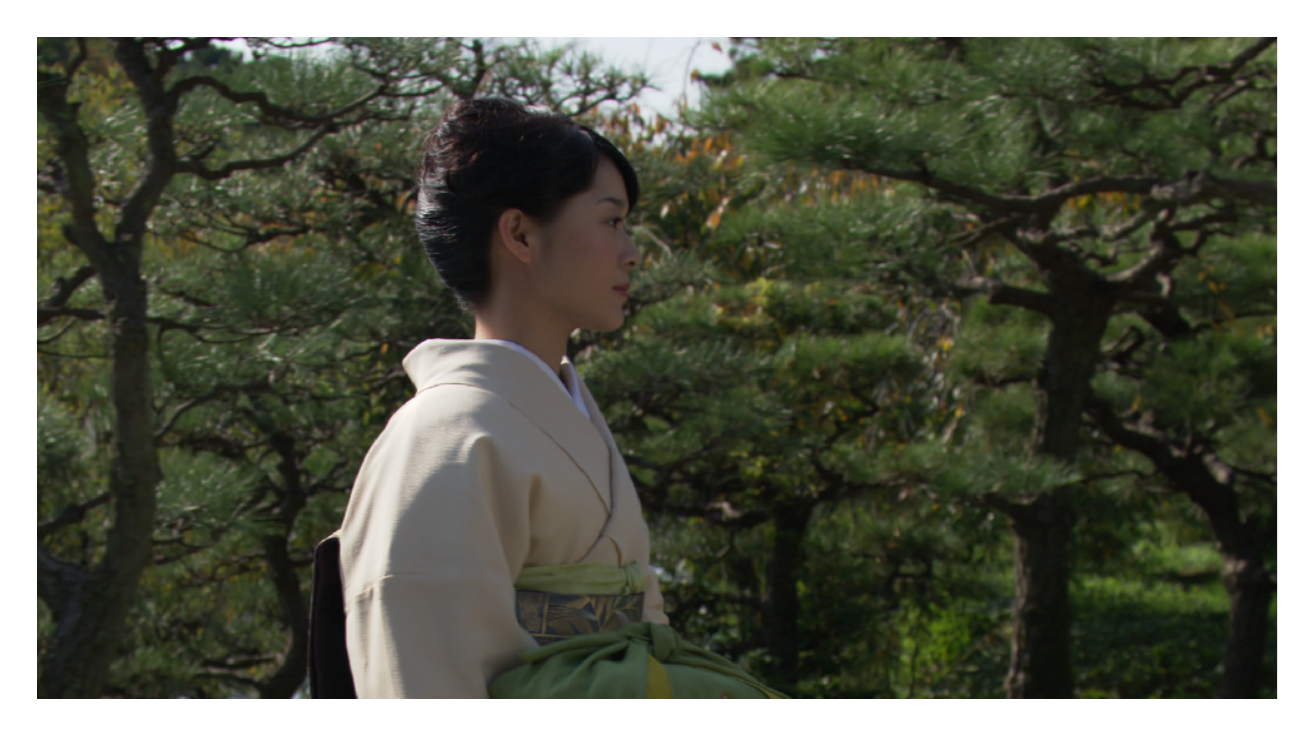}
    }
  \subfigure[]{
    \includegraphics[width=.17\linewidth]{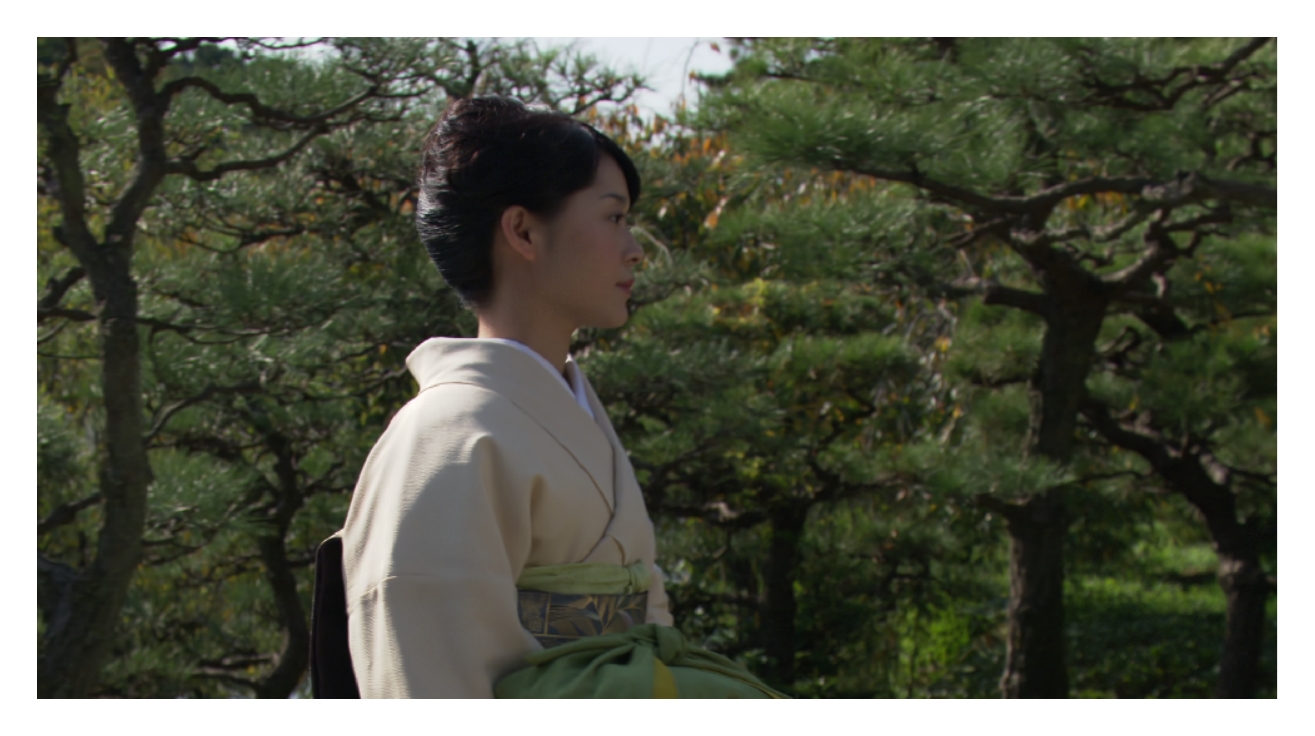}}
  \subfigure[]{
    \includegraphics[width=.17\linewidth]{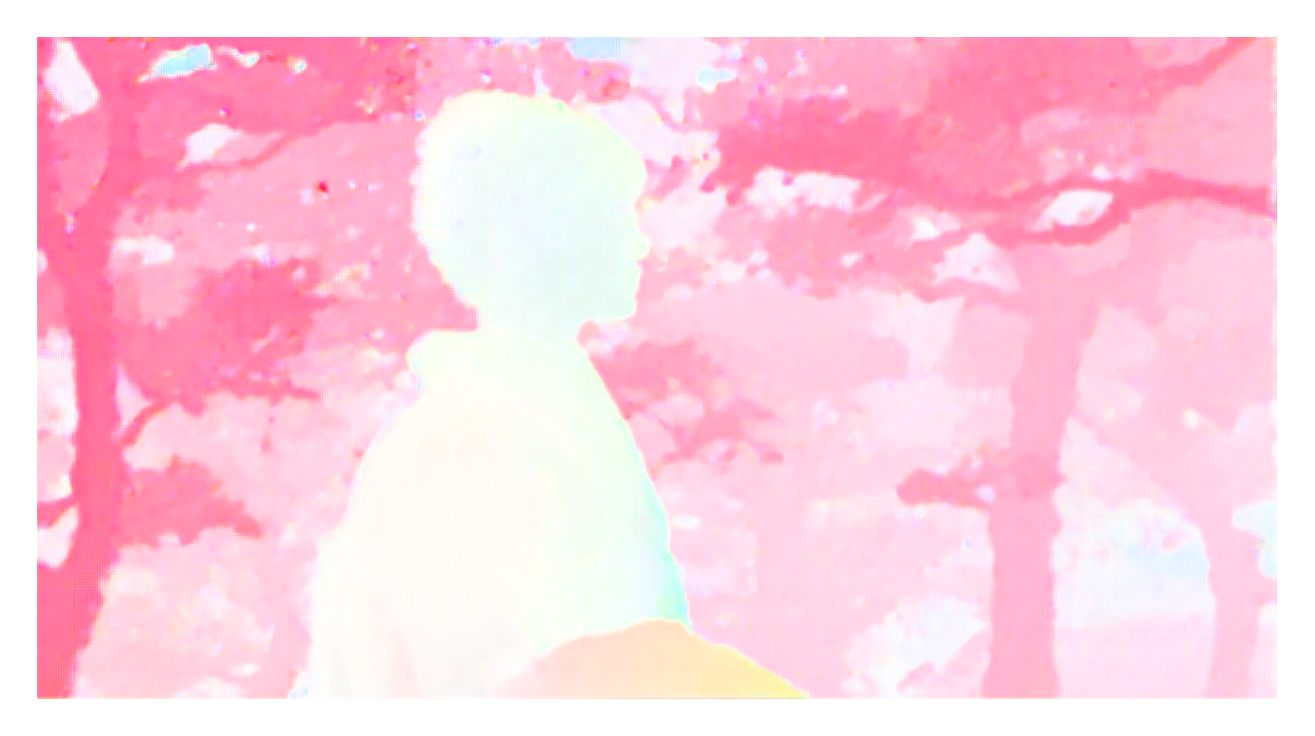}}
  \subfigure[]{
    \includegraphics[width=.17\linewidth]{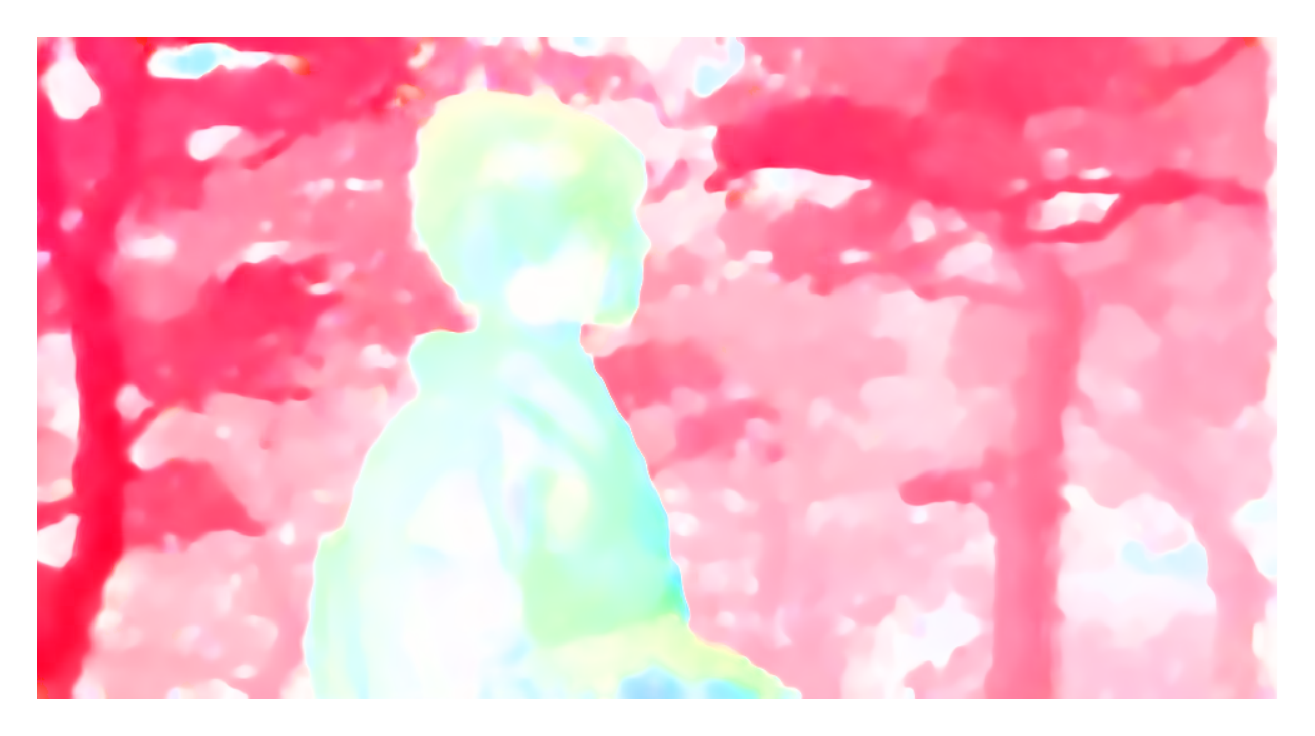}}
  \subfigure[]{
    \includegraphics[width=.17\linewidth]{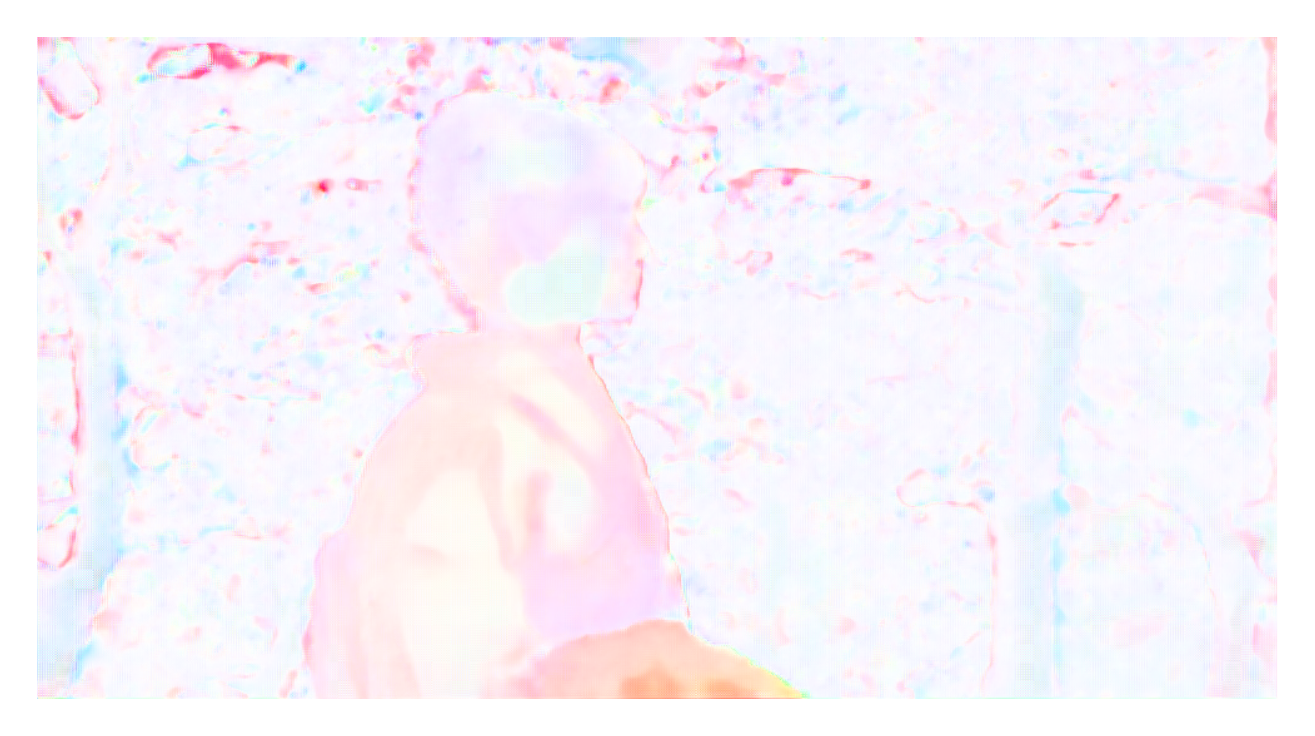}}
  \caption{Visualized results of compressing the Kimono sequence using \texttt{Add MAMVP-Net} model with $\lambda =16$. (a) The reference frame $\hat{x}_{5}$. (b) The original frame $x_{6}$. (c) The original MV $v_{6}$. (d) The predicted MV $\bar{v}_{6}$. (e) The MVD $d_{6}$.}
  \label{fig_mvp} 
\end{figure*}
\begin{table*}
\caption{Average running time per frame of using our different models for a 320$\times$256 sequence.}
\label{Time}
\center
\begin{tabular}{c|c|c|c|c|c}
\hline
Model          & \texttt{Our Baseline}    & \texttt{Add MAMVP-Net}  & \texttt{Add MVRefine-Net}   & \texttt{Add MMC-Net} & \texttt{Proposed} \\
\hline
Encoding Time        & 0.25s                   &0.31s	           &0.34s	          &0.35s	& 0.37s   \\
\hline
Decoding Time        & 0.05s                    &0.11s	           &0.14s	          &0.15s	& 0.17s   \\
\hline
\end{tabular}
\end{table*}
\begin{figure}
\begin{center}
\subfigure[]{
    \includegraphics[width=.40\linewidth]{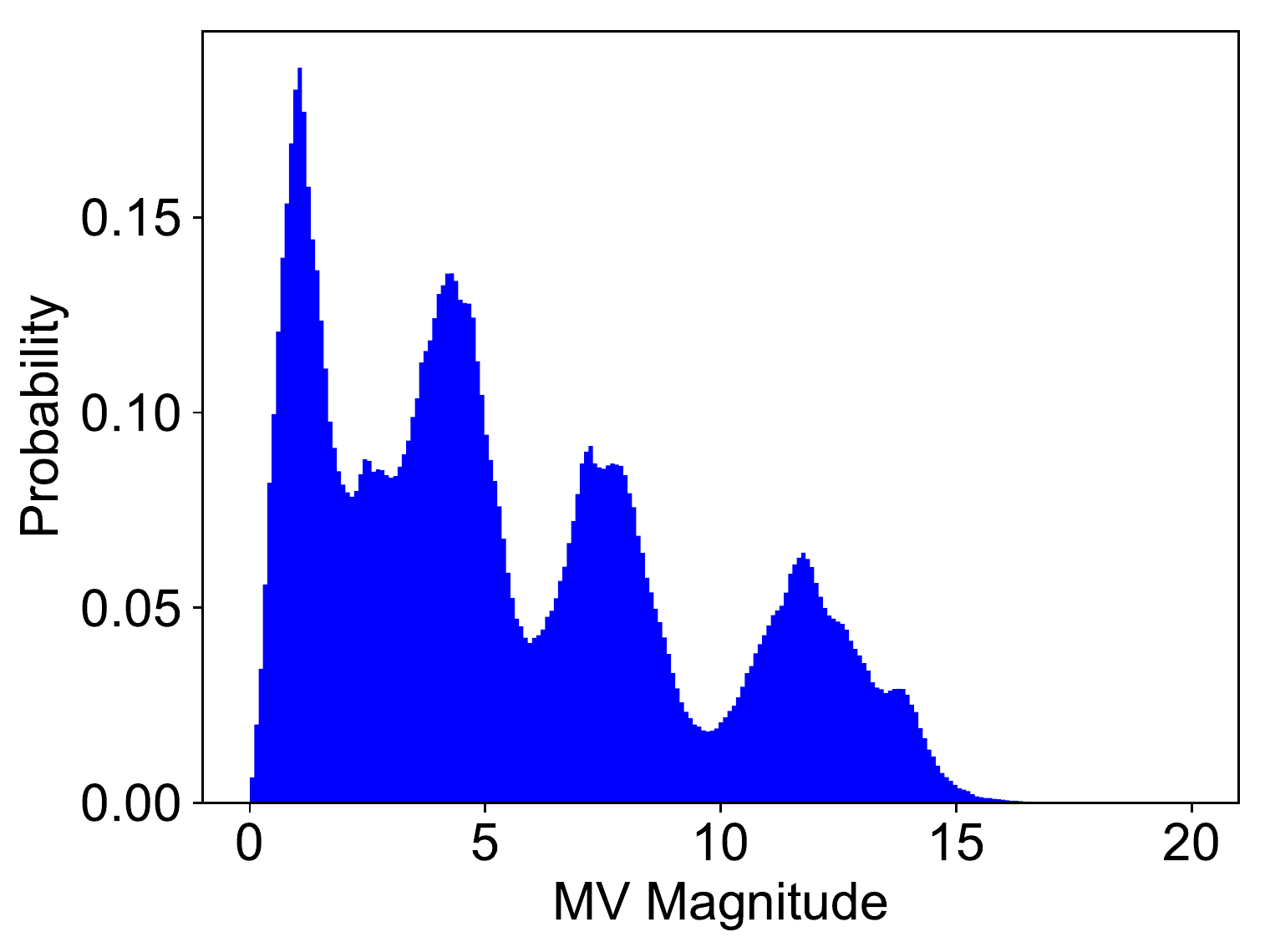}}
  \subfigure[]{
    \includegraphics[width=.40\linewidth]{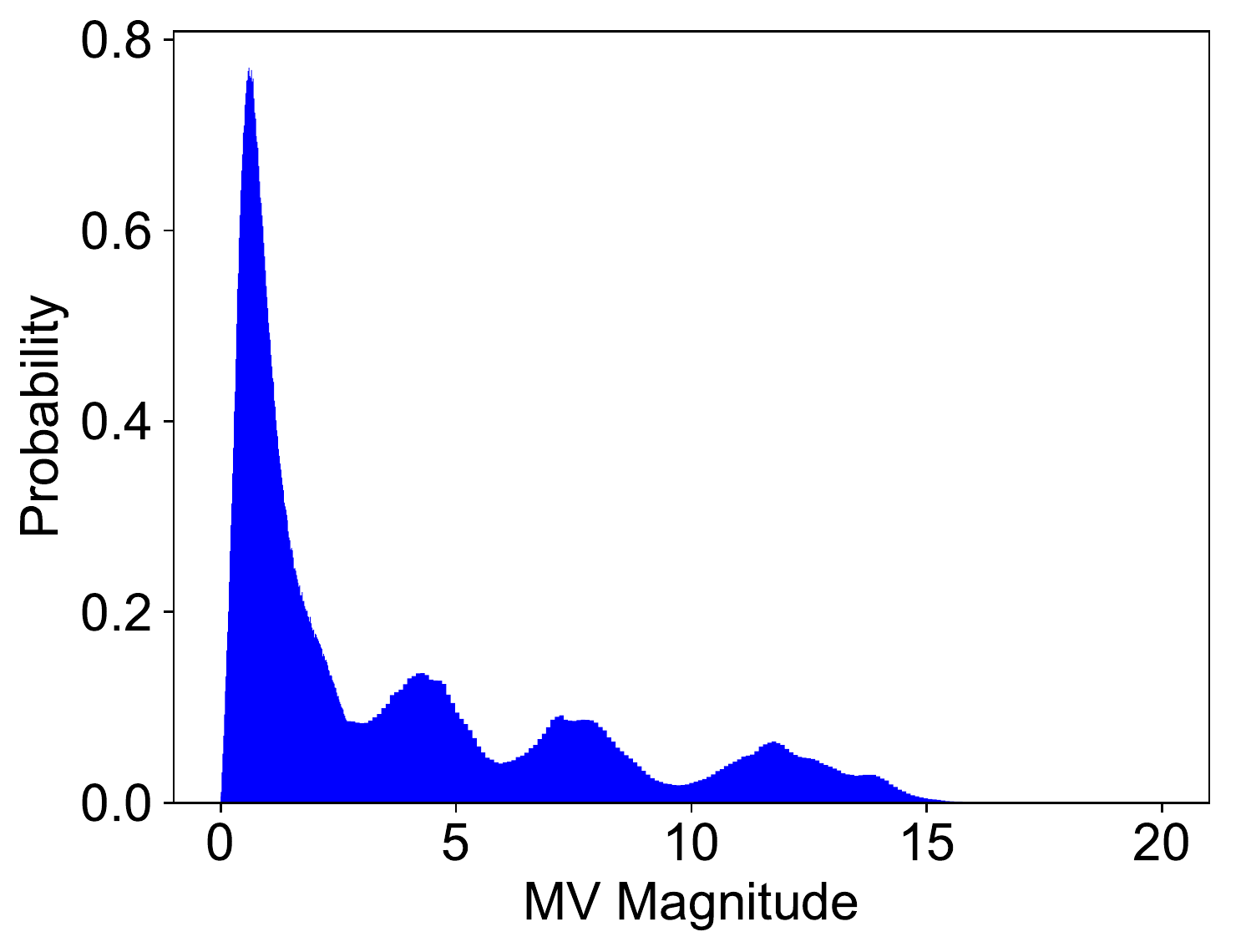}}
\end{center}
   \caption{The distribution of MV magnitude. (a) The MV of Fig.\ \ref{fig_mvp} (c). (b) The MVD of Fig.\ \ref{fig_mvp} (e).}
\label{fig:Flow_dis}
\end{figure}
To evaluate its effectiveness, we perform a comparison experiment. The anchor is the network containing the ME-Net, the MC-Net with only one reference frame, and the MV and residual encoder-decoder networks. Here, the MC-Net with only one reference frame is almost identical to the MMC-Net shown in Fig.\ \ref{fig:MCNet}, except for removing $f_{\hat{x}_{t-4}}^{w}$, $f_{\hat{x}_{t-3}}^{w}$, $f_{\hat{x}_{t-2}}^{w}$ from the inputs. This anchor is denoted by \texttt{Our Baseline} (the green curve in Fig.\ \ref{fig:Ablation}). The tested network is constructed by adding the MAMVP-Net to \texttt{Our Baseline}, and is denoted by \texttt{Add MAMVP-Net} (the red curve in Fig.\ \ref{fig:Ablation}). It can be observed that the MAMVP-Net improves the compression efficiency significantly, achieving about $0.5\sim0.7$ dB gain at the same bpp.
In Fig.\ \ref{fig_mvp}, we visualize the intermediate results when compressing the Kimono sequence using \texttt{Add MAMVP-Net} model. Fig.\ \ref{fig:Flow_dis} shows the corresponding probability distributions of MV magnitudes for $v_{6}$ and $d_{6}$. It is observed that the magnitude of MV to be encoded is greatly reduced by using our MAMVP-Net. Quantitatively, it needs 0.042bpp for encoding the original MV $v_6$ using \texttt{Our Baseline} model, while it needs 0.027bpp for encoding the MVD $d_6$ using \texttt{Add MAMVP-Net} model. Therefore, our MAMVP-Net can largely reduce the bits for encoding MV and thus improve the compression efficiency. More ablation study results can be found in the supplementary.

{\bf MV Refinement Network.}
To evaluate the effectiveness, we perform another experiment by adding the MV Refine-Net to \texttt{Add MAMVP-Net}, leading to \texttt{Add MVRefine-Net} (the cyan curve in Fig.\ \ref{fig:Ablation}). Compared with the compression results of \texttt{Add MAMVP-Net}, at the same bpp, the MV Refine-Net achieves a compression gain of about 0.15dB at high bit rates and about 0.4dB at low bit rates. This is understandable as the compression error is more severe when the bit rate is lower. In addition, to evaluate the effectiveness of introducing $\hat{x}_{t-1}$ into the MV Refine-Net, we perform an experiment by removing $f_{\hat{x}_{t-1}}$ from the inputs of the MV Refine-Net (denoted by \texttt{Add MVRefine-Net-0}, the black curve in Fig.\ \ref{fig:Ablation}). We can observe that feeding $\hat{x}_{t-1}$ into the MV Refine-Net provides about 0.1dB gain consistently. Visual results of the MV Refine-Net can be found in the supplementary.

{\bf Motion Compensation Network with Multiple Reference Frames.}
To verify the effectiveness, we perform an experiment by replacing the MC-Net (with only one reference frame) in \texttt{Add MVRefine-Net} with the proposed MMC-Net using multiple reference frames (denoted by \texttt{Add MMC-Net}, the magenta curve in Fig.\ \ref{fig:Ablation}). We can observe that using multiple reference frames in MMC-Net provides about $0.1\sim0.25$dB gain. Visual results of the MMC-Net can be found in the supplementary.

{\bf Residual Refinement Network.}
We conduct another experiment to evaluate its effectiveness by adding the Residual Refine-Net to \texttt{Add MMC-Net} (denoted by \texttt{Proposed}, the blue curve in Fig.\ \ref{fig:Ablation}). We observe that the Residual Refine-Net provides about 0.3dB gain at low bit rates and about 0.2dB gain at high bit rates. Similar to MV Refine-Net, the gain of Residual Refine-Net is higher at lower bit rates because of more compression error. Visual results of the Residual Refine-Net can be found in the supplementary.

{\bf Step-by-step Training Strategy.}
To verify the effectiveness, we perform an experiment by training the \texttt{Proposed} model from scratch except the ME-Net initialized by the pre-trained model in \cite{ilg2017flownet} (denoted by \texttt{Scratch}, the yellow curve in Fig.\ \ref{fig:Ablation}). We can observe that the compression results are very bad. Quantitatively, when compressing the Kimono sequence using \texttt{Scratch} model with $\lambda =16$, the bitrates are very unbalanced: 0.0002bpp for MVD and 0.2431bpp for residual. Our step-by-step training strategy can overcome this.

{\bf Encoding and Decoding Time.}
We use a single Titan Xp GPU to test the inference speed of our different models. The running time is presented in Table \ref{Time}. We can observe that the MAMVP-Net increases more encoding/decoding time than the other newly added modules. For a 352$\times$256 sequence, the overall encoding (resp. decoding) speed of our \texttt{Proposed} model is 2.7fps (resp. 5.9fps). It requires our future work to optimize the network structure for computational efficiency to achieve real-time decoding.

\vspace{-0.18cm}
\section{Conclusion}

In this paper, we have proposed an end-to-end learned video compression scheme for low-latency scenarios. Our scheme can effectively remove temporal redundancy by utilizing multiple reference frames for both motion compensation and motion vector prediction. We also introduce the MV and residual refinement modules to compensate for the compression error and to enhance the reconstruction quality. All the modules in our scheme are jointly optimized by using a single rate-distortion loss function, together with a step-by-step training strategy. Experimental results show that our method outperforms the existing learned video compression methods for low-latency mode. In the future, we anticipate that advanced entropy coding model can further boost the compression efficiency.

{\small
\bibliographystyle{ieee_fullname}
\bibliography{egbib}
}

\appendix

\section{Proposed Method}

\subsection{Details of Our MV Refinement Network}
\begin{figure}[t]
\begin{center}
\includegraphics[width=.9\linewidth]{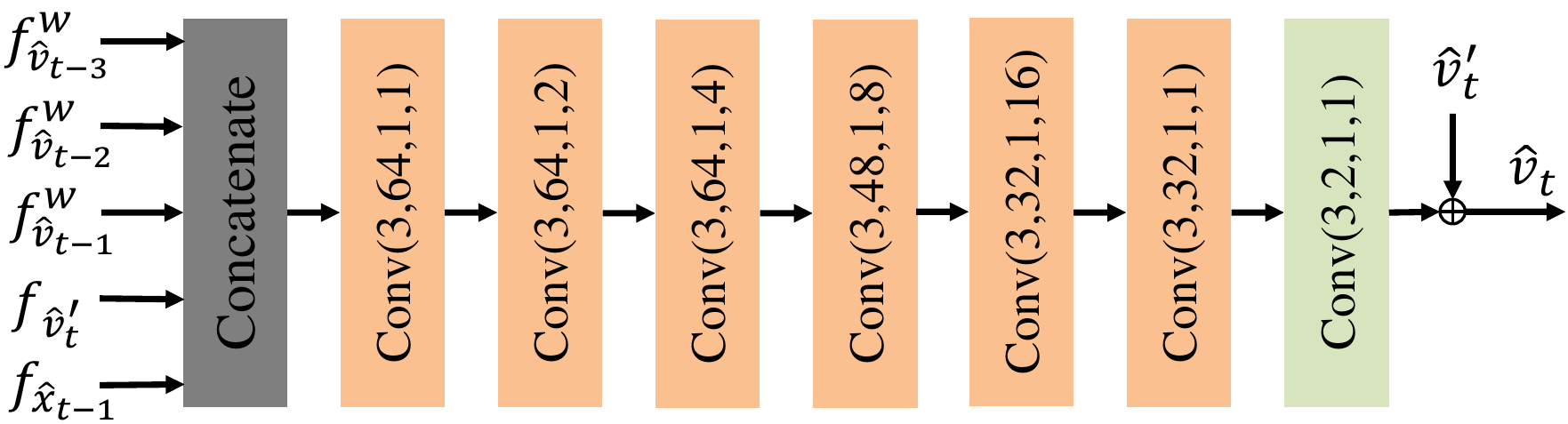}
\end{center}
   \caption{The MV refinement network. Conv(3,64,1,1) represents the convolutional layer with the kernel size of 3$\times$3, the output channel of 64, the stride of 1, and the dilation constant of 1. Each convolutional layer is followed by a leaky ReLU except the last layer (indicated by green).}
\label{fig:MVRNet}
\end{figure}
The architecture of our MV refinement network is presented in Fig.~\ref{fig:MVRNet}. We first use a two-layer CNN to extract the features of $\hat{v}_{t-3}$, $\hat{v}_{t-2}$, $\hat{v}_{t-1}$, $\hat{v}_{t}'$, and $\hat{x}_{t-1}$, respectively. And then, the features of $\hat{v}_{t-3}$, $\hat{v}_{t-2}$ and $\hat{v}_{t-1}$ are warped towards $v_{t}$ with the help of $\hat{v}_{t}'$,
\begin{equation}\label{warp_mvr}
\begin{split}
  \hat{v}^{w}_{t-k} &= Warp(\hat{v}_{t-k},\hat{v}_{t}'+\sum_{l=1}^{k-1}\hat{v}^{w}_{t-l}), k=1,2\\
  f_{\hat{v}_{t-i}}^{w} &= Warp(f_{\hat{v}_{t-i}},\hat{v}_{t}'+\sum_{k=1}^{i-1}\hat{v}^{w}_{t-k}), i=1,2,3 \\
\end{split}
\end{equation}
where $\hat{v}^{w}_{t-k}$ is the warped version of $\hat{v}_{t-k}$ towards $\hat{v}_{t}'$. Finally, the warped features, and the features of $\hat{v}_{t}'$ and $\hat{x}_{t-1}$ are fed into a dilated convolution-based network, which can capture larger receptive field, to obtain the final reconstructed MV,
\begin{equation}\label{refine_mvr}
  \hat{v}_{t} = H_{mvr}(f_{\hat{v}_{t-3}}^{w},f_{\hat{v}_{t-2}}^{w},f_{\hat{v}_{t-1}}^{w},f_{\hat{v}_{t}'},f_{\hat{x}_{t-1}})+\hat{v}_{t}'
\end{equation}
where $H_{mvr}$ denotes the function of the network.

\subsection{Details of Our Residual Refinement Network}
\begin{figure}[t]
\begin{center}
\includegraphics[width=.9\linewidth]{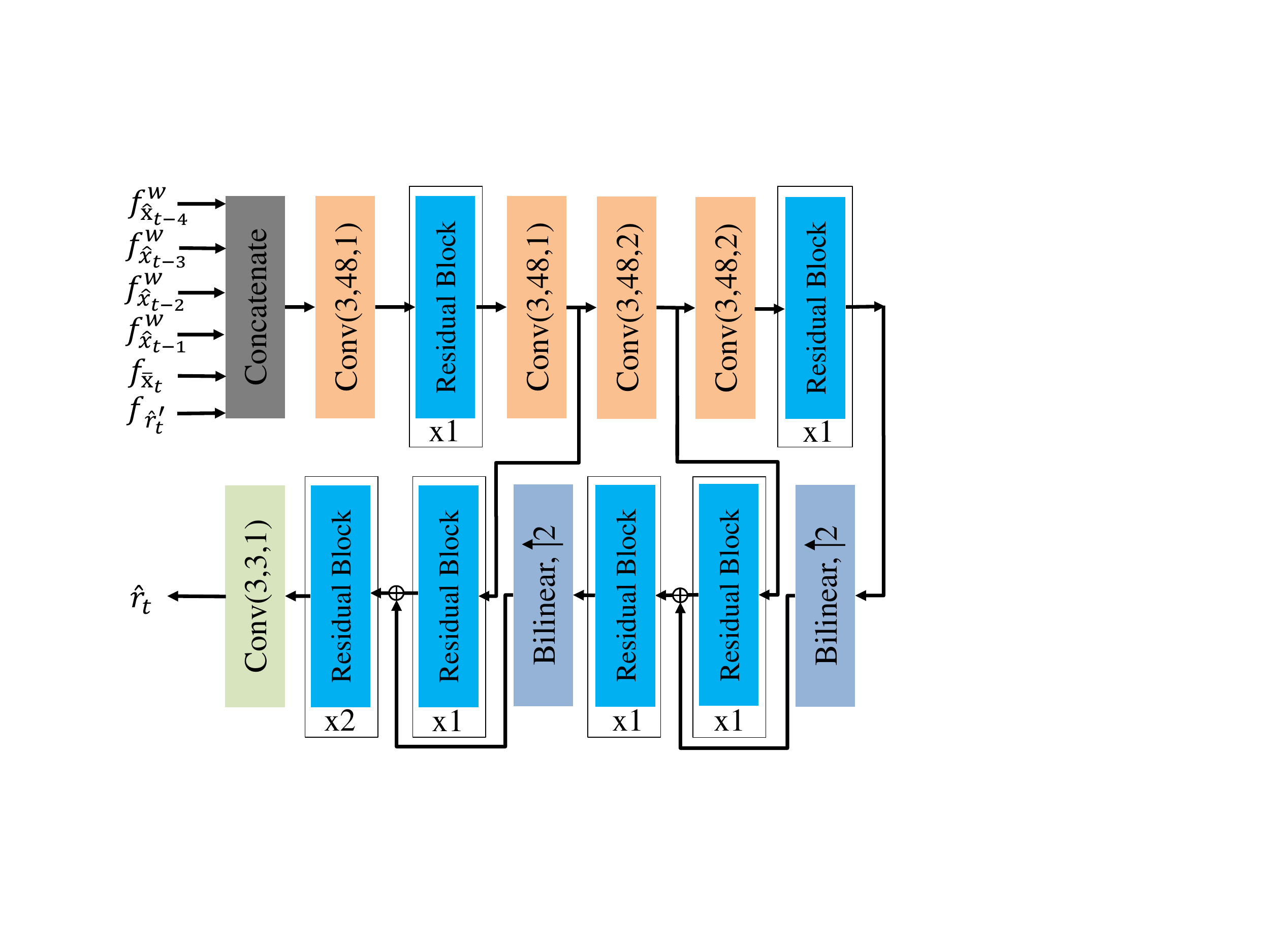}
\end{center}
   \caption{The residual refinement network. Each convolutional layer outside residual blocks is followed by a leaky ReLU except the last layer (indicated by green). Each residual block consists of two convolutional layers, which are configured as follows: kernel size is 3$\times$3, output channel number is 48, the first layer has ReLU.}
\label{fig:RRNet}
\end{figure}
Fig.~\ref{fig:RRNet} shows the architecture of our residual refinement network. First, we use a two-layer CNN to extract the features of $\hat{x}_{t-4}$, $\hat{x}_{t-3}$, $\hat{x}_{t-2}$, and $\hat{x}_{t-1}$ and warp them towards the current frame. This warping operation is the same with Eq. (4) in the paper. Then, the warped features and the features of $\bar{x}_{t}$ and $\hat{r}_{t}'$ are fed into a CNN, which is based on the U-Net structure \cite{ronneberger2015u} and integrates multiple residual blocks, to obtain the refined residual $\hat{r}_{t}$,
\begin{equation}\label{DRP-equation}
  \hat{r}_{t} = H_{res}(f_{\hat{x}_{t-4}}^{w}, f_{\hat{x}_{t-3}}^{w}, f_{\hat{x}_{t-2}}^{w}, f_{\hat{x}_{t-1}}^{w}, f_{\bar{x}_{t}}, f_{\hat{r}_{t}'})
\end{equation}
 where $H_{res}$ represents the function of the network.
\section{Experiments}
\subsection{Ablation Study of Our MAMVP-Net}
\begin{table}
\centering
\caption{Bit-rates (bpp) and reconstruction quality (PSNR) for ablation study of the MAMVP-Net}\label{mamvp_ablation}
\resizebox{\columnwidth}{!}{
\begin{tabular}{c|c|c|c|c}
\hline
Network &single-scale   &single-scale	&multi-scale    &multi-scale   	                        	             \\
&w/o alignment &w/ alignment &w/o alignment & w/ alignment \\
\hline
bpp                 &0.297      &0.290      &0.287   &0.285                             	             \\
\hline
PSNR (dB)                  &31.250      &31.198      &31.196 &31.290                                         \\
\hline
\end{tabular}
}
\end{table}
To verify the effectiveness of the components in MAMVP-Net, we conduct experiments to compare the proposed MAMVP-Net (denoted by multi-scale w/ alignment) with its simplified versions: (1) single-scale w/o alignment, (2) single-scale w/ alignment, (3) multi-scale w/o alignment. These models are tested on HEVC Class D dataset and the reconstruction quality and bit-rates are shown in Table \ref{mamvp_ablation}. It can be observed that the proposed MAMVP-Net achieves the highest reconstruction quality with the lowest bit-rates.

\subsection{Visual Results of Our MV Refine-Net}
\begin{figure}
\begin{center}
    \includegraphics[width=.80\linewidth]{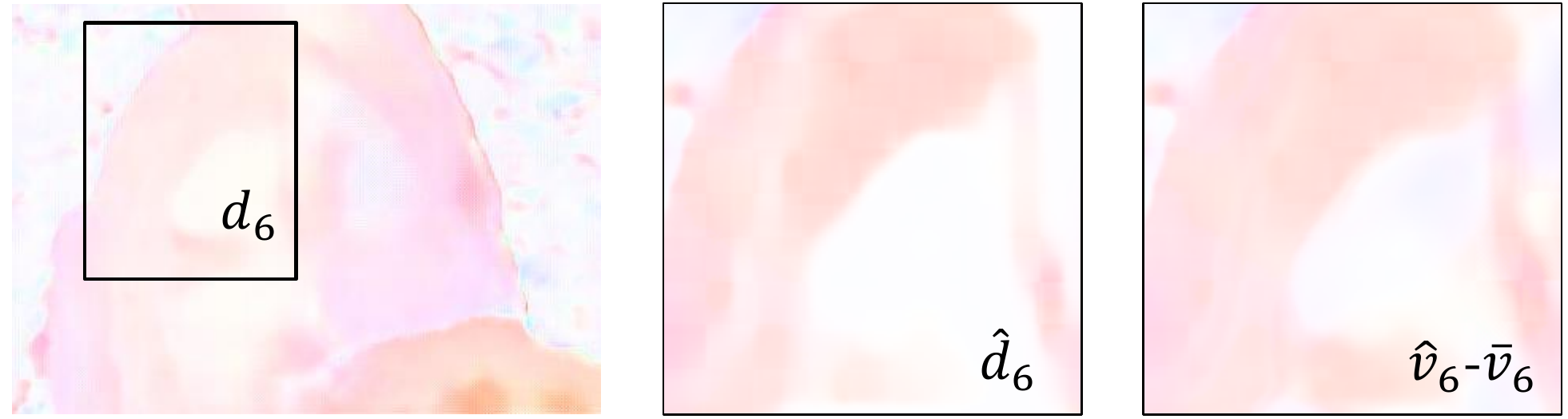}
\end{center}
   \caption{Visualized results of compressing the Kimono sequence using \texttt{Add MVRefine-Net} model. From left to right: the original MVD $d_{6}$, the decoded MVD $\hat{d}_{6}$, and the refined MVD, \ie $\hat{v}_{6}-\bar{v}_{6}$.}
\label{fig:MVR_Vis}
\vspace{-0.2cm}
\end{figure}
In Fig.\ \ref{fig:MVR_Vis}, we visualize the original MVD $d_{6}$, the decoded MVD $\hat{d}_{6}$, and the MVD after refinement, \ie $\hat{v}_{6}-\bar{v}_{6}$, when compressing the Kimono sequence using \texttt{Add MVRefine-Net} model. After compression, there are more zeros in $\hat{d}_{6}$ than $d_{6}$ due to the bit rate constraint. Our MV Refine-Net can restore some non-zero MVDs and thus improve the accuracy.
\subsection{Visual Results of Our MMC-Net}
\begin{figure*}
  \centering
  \subfigure[]{
    \includegraphics[width=.3\linewidth]{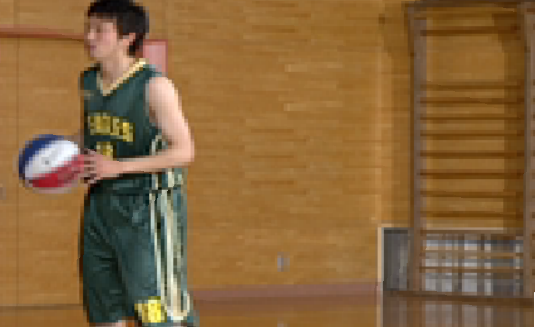}
    }
  \subfigure[]{
    \includegraphics[width=.3\linewidth]{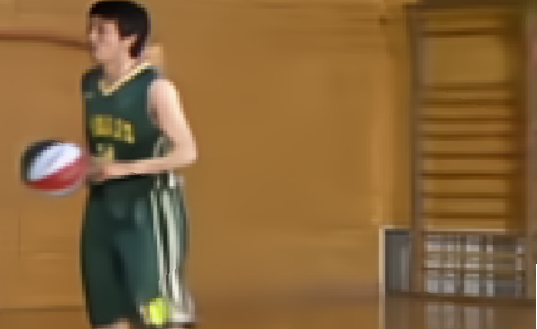}}
  \subfigure[]{
    \includegraphics[width=.3\linewidth]{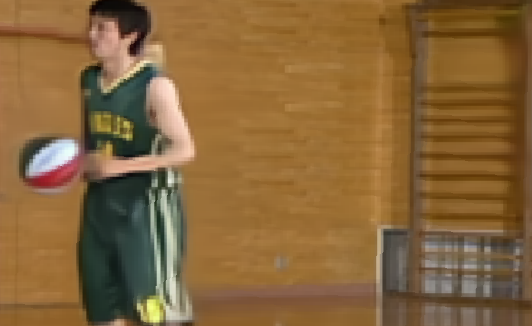}}
  \caption{Visualized results of compressing the BasketballPass sequence. (a) The original frame $x_{9}$. (b) The predicted frame $\bar{x}_{9}$ obtained by \texttt{Add MVRefine-Net} model with $\lambda=$ 64. (c) The predicted frame $\bar{x}_{9}$ obtained by \texttt{Add MMC-Net} model with $\lambda=$ 64. There are much more details in (c) than (b).}
  \label{fig:MMC_Vis} 
\vspace{-0.3cm}
\end{figure*}
 In Fig.~\ref{fig:MMC_Vis}, we visualize the original frame $x_{9}$ (a), the predicted frame $\bar{x}_{9}$ obtained by \texttt{Add MVRefine-Net} model with $\lambda=$ 64 (b), and the predicted frame $\bar{x}_{9}$ obtained by \texttt{Add MMC-Net} model with $\lambda=$ 64 (c), when compressing the BasketballPass sequence. We can observe that the image in Fig.~\ref{fig:MMC_Vis} (b) is much more smooth than (c), \eg in the area of the wall. Quantitatively, the PSNR of the predicted frame in Fig.~\ref{fig:MMC_Vis} (c) is 31.97dB, while the PSNR of the predicted frame in Fig.~\ref{fig:MMC_Vis} (b) is 31.42dB. Therefore, our MMC-Net can obtain a more accurate prediction with more details by using multiple reference frames.
\subsection{Visual Results of Our Residual Refine-Net}
\begin{figure}
\begin{center}
    \includegraphics[width=.80\linewidth]{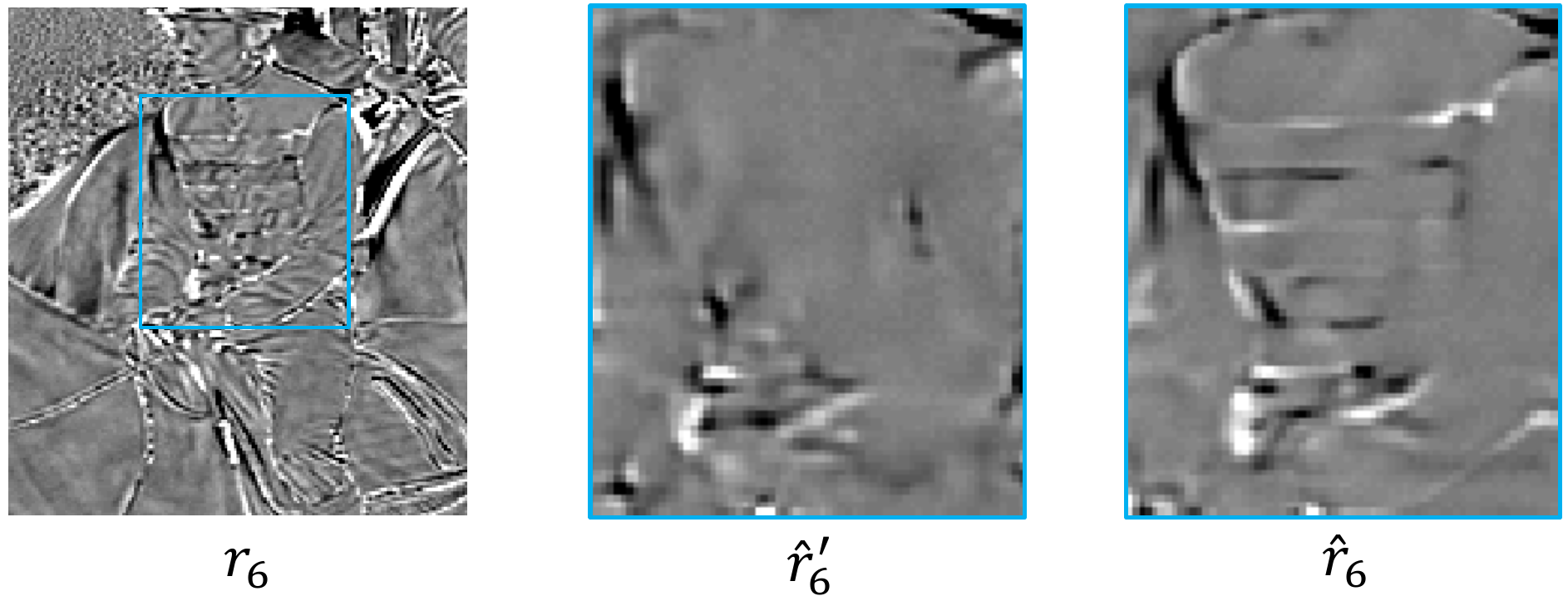}
\end{center}
   \caption{Visualized results of compressing the RaceHorses sequence using \texttt{Proposed} model. From left to right: the original residual $r_{6}$, the decoded residual $\hat{r}_{6}'$, and the refined residual $\hat{r}_{6}$.}
\label{fig:RR_Vis}
\vspace{-0.5cm}
\end{figure}
In Fig.\ \ref{fig:RR_Vis}, we visualize the original residual $r_{6}$, the decoded residual $\hat{r}_{6}'$, and the refined residual $\hat{r}_{6}$, when compressing the RaceHorses sequence using \texttt{Proposed} model. We can observe that $\hat{r}_{6}'$ is much more smooth than $r_6$ due to the rate constraint. Our Residual Refine-Net can restore some image details and thus improve the reconstruction quality.

\subsection{Compression Performance on the HEVC Class C and E Datasets}
\begin{figure*}
  \centering
  \subfigure[]{
    \includegraphics[width=3.0in]{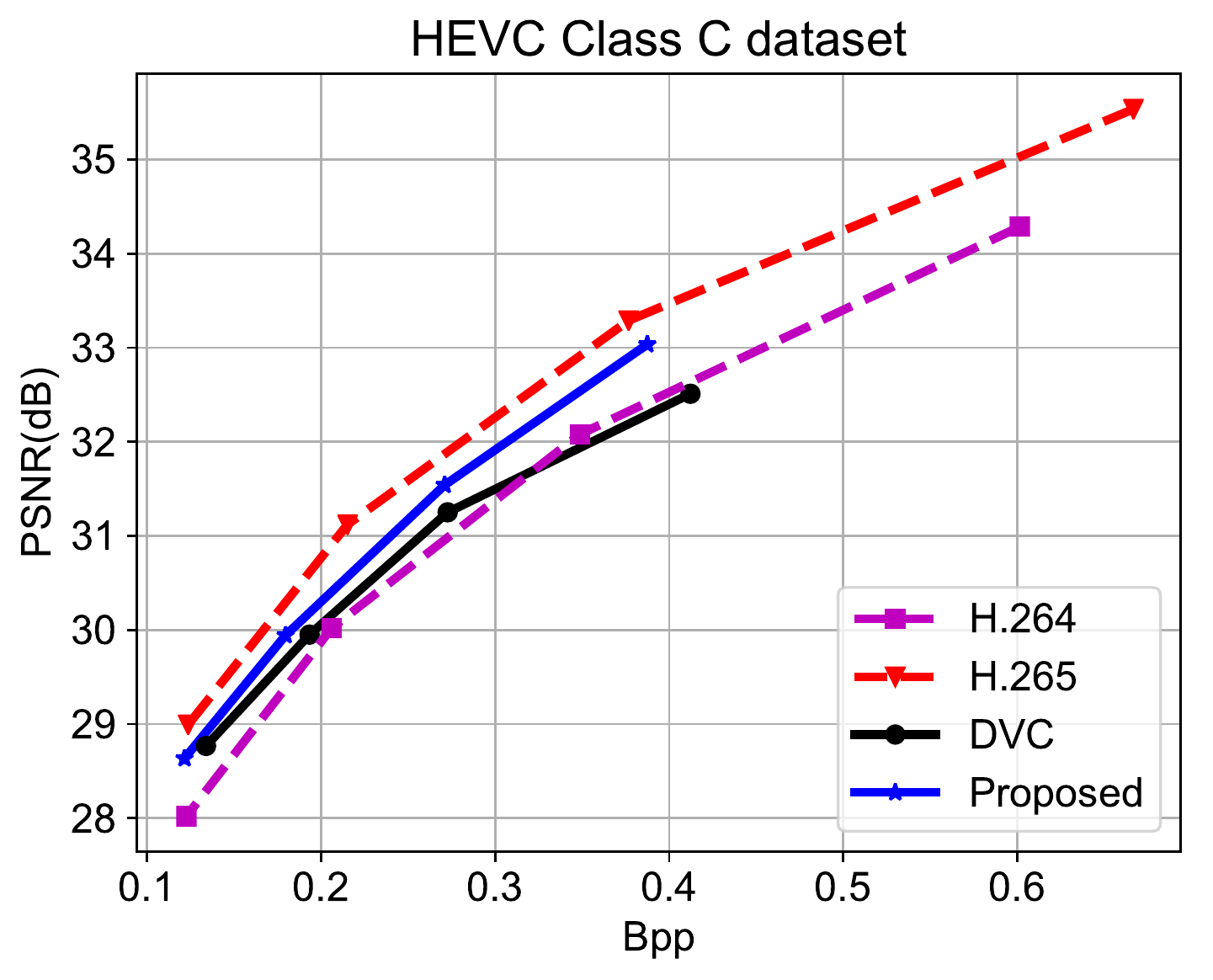}}
  \subfigure[]{
    \includegraphics[width=3.0in]{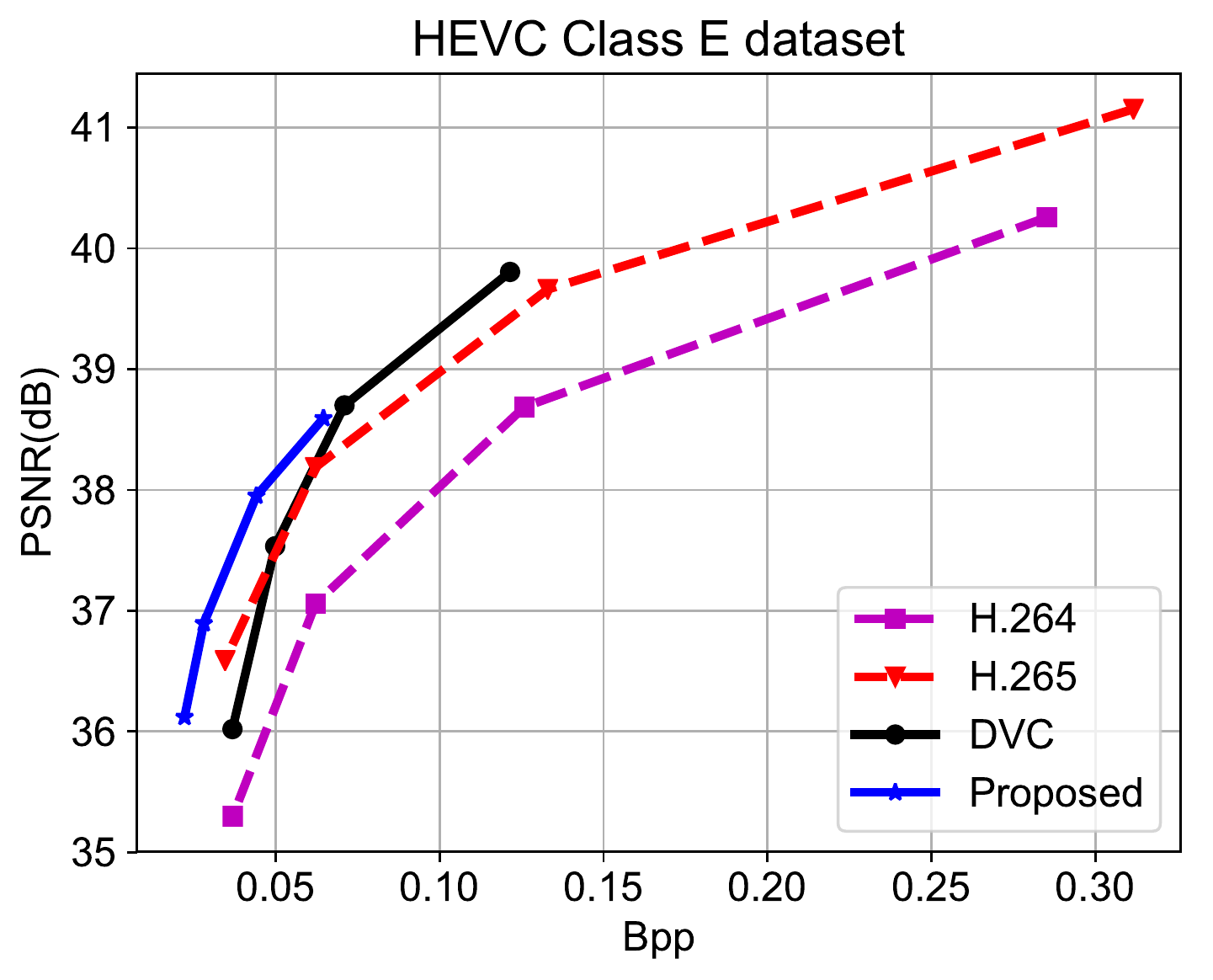}}
  \subfigure[]{
    \includegraphics[width=3.0in]{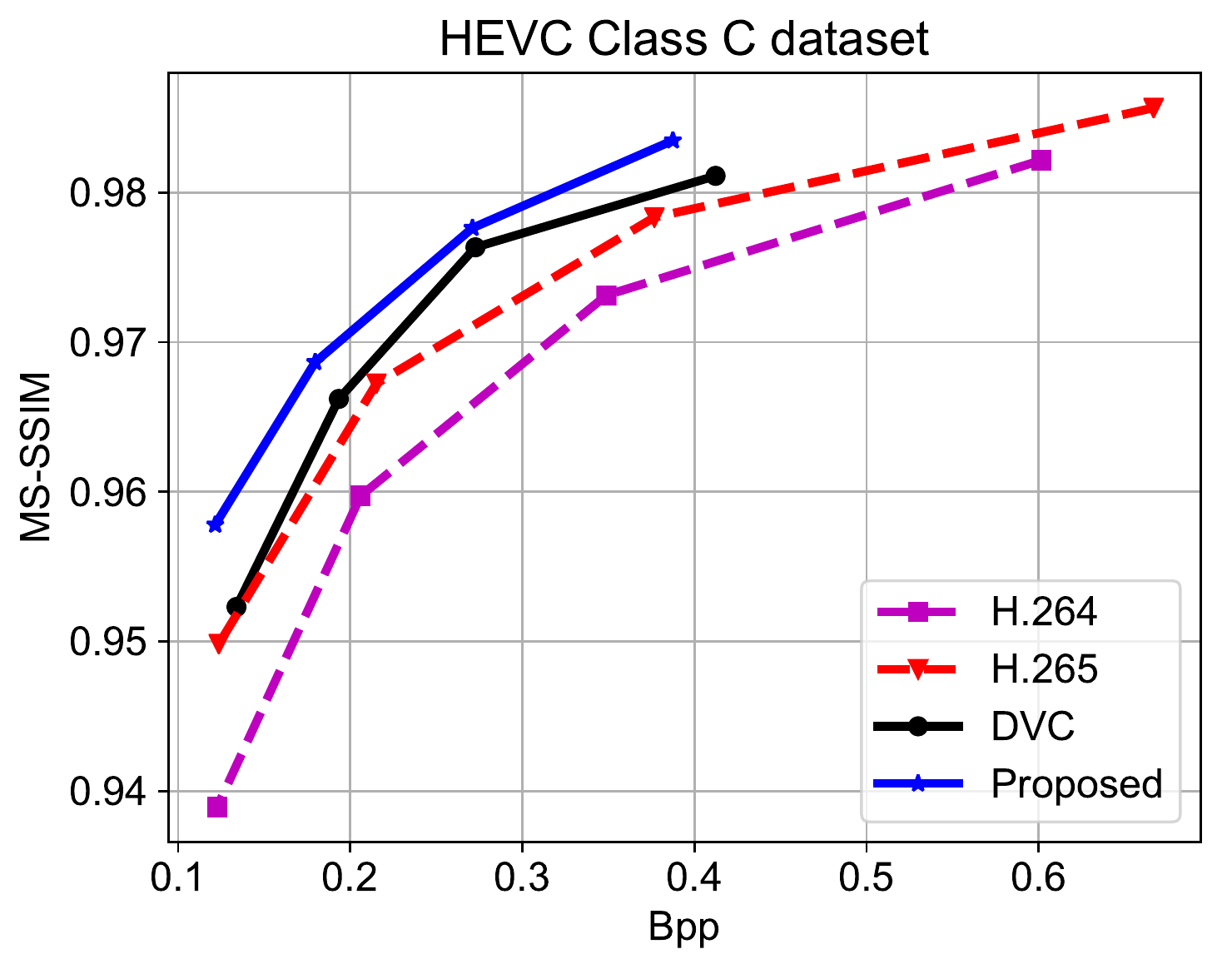}}
  \subfigure[]{
    \includegraphics[width=3.0in]{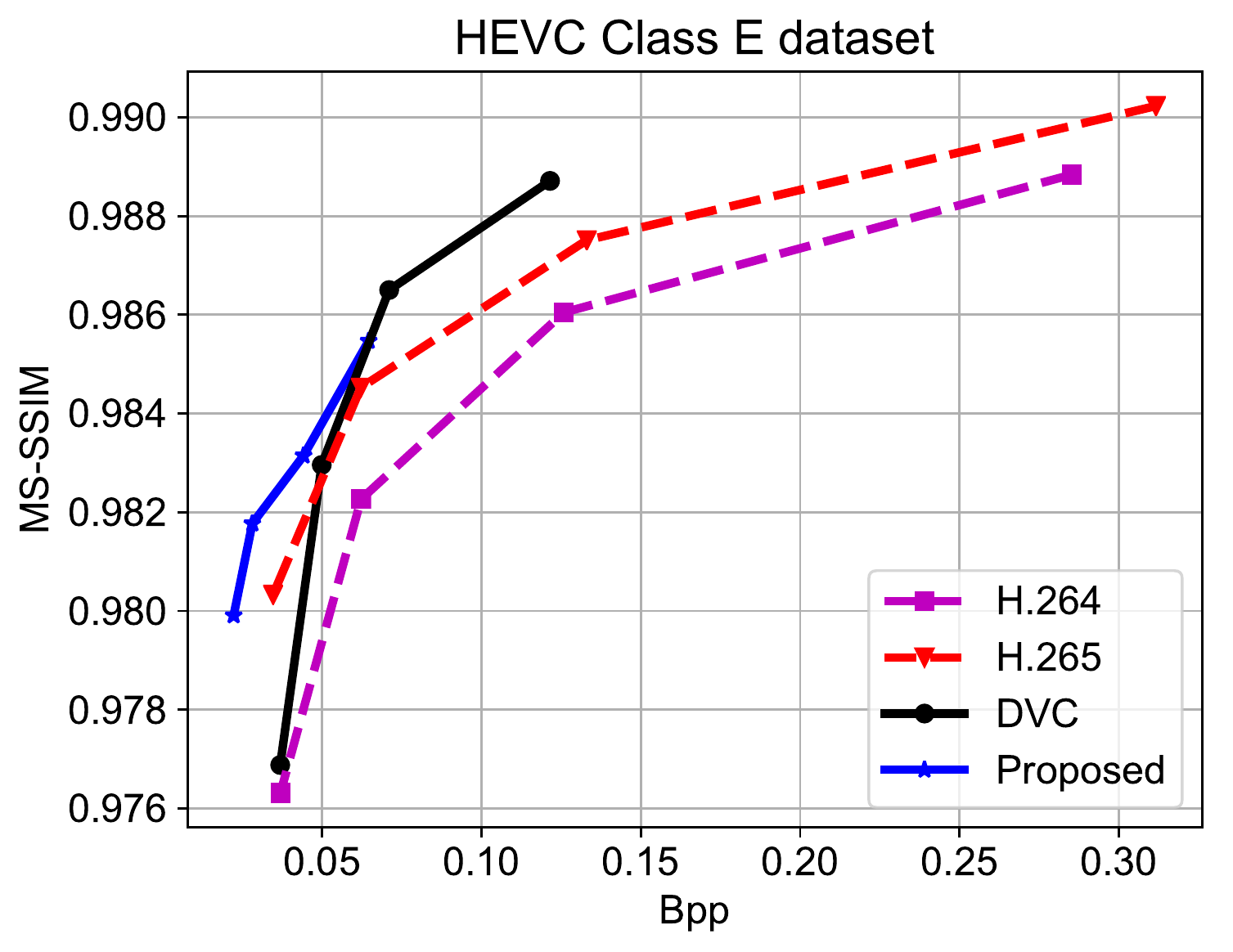}}
  \caption{Compression results of H.264, H.265, DVC \cite{lu2018dvc}, and the proposed method on the HEVC Class C and E datasets. The results of H.264 and H.265 are cited from \cite{lu2018dvc}.}
  \label{fig_RD Curve} 
\end{figure*}
We provide the compression results on the HEVC Class B and D datasets in the paper. In Fig.~\ref{fig_RD Curve}, we also present the compression results on the HEVC Class C and E datasets using H.264, H.265, DVC \cite{lu2018dvc}, and the proposed method. It can be observed that our method outperforms DVC \cite{lu2018dvc} by a large margin. When compared with H.265, our method achieves on par or better compression performance in PSNR and MS-SSIM.

\subsection{Comparison with Other Learned Video Compression Methods}
\begin{figure*}
  \centering
  \subfigure[]{
    \includegraphics[width=3.0in]{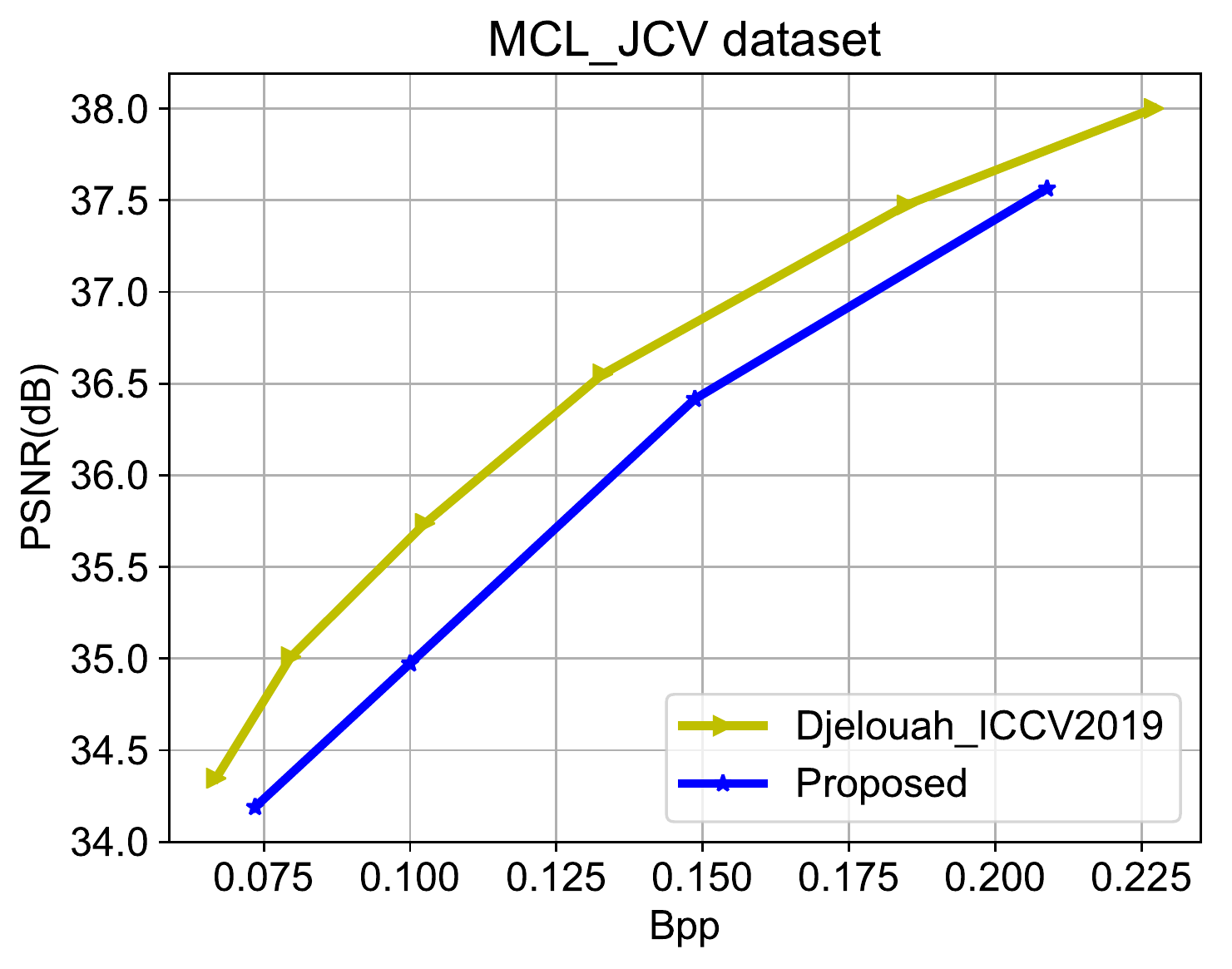}}
  \subfigure[]{
    \includegraphics[width=3.0in]{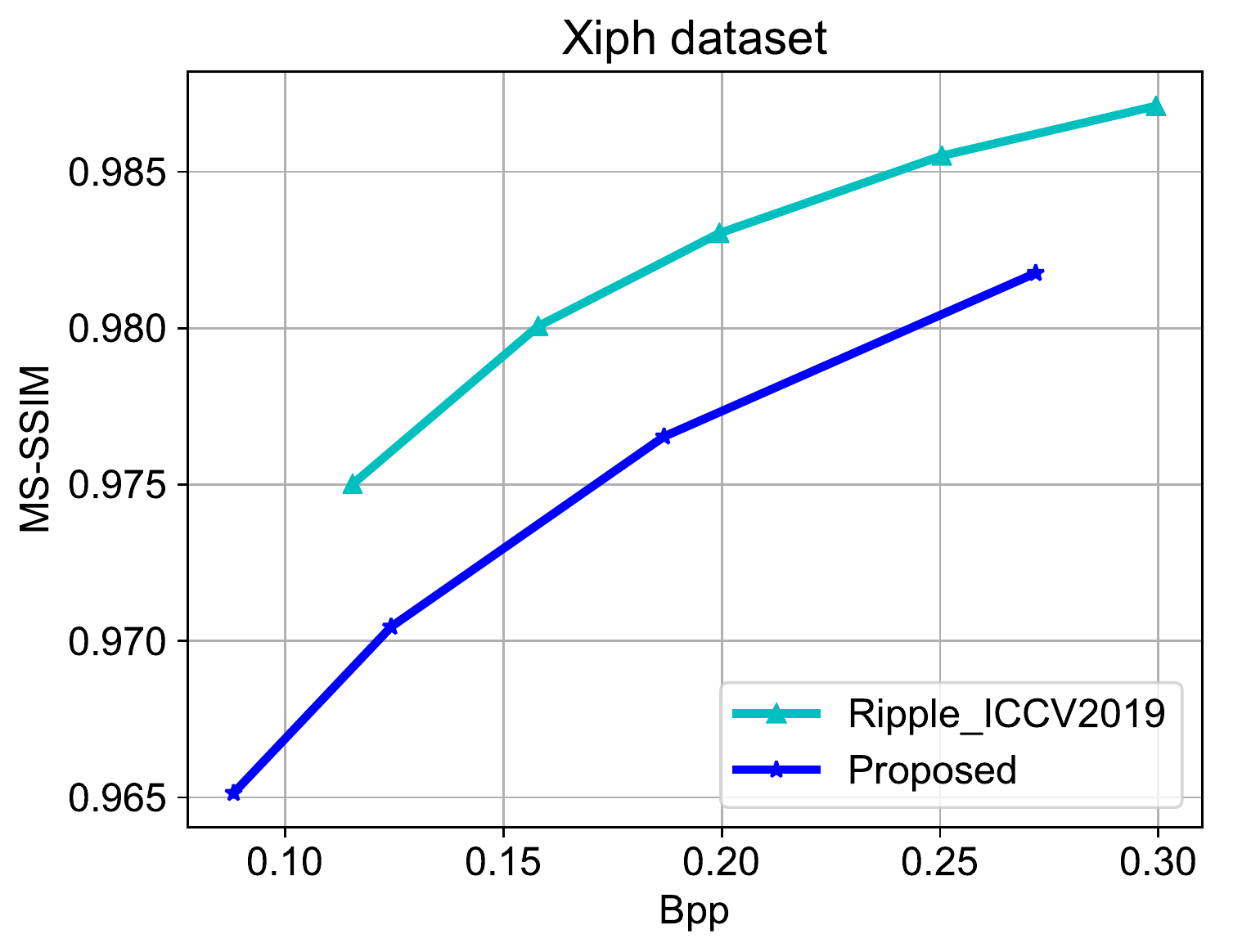}}
  \caption{Compression results of Djelouah\_ICCV2019 \cite{Djelouah_2019_ICCV}, Rippel\_ICCV2019 \cite{Rippel_2019_ICCV}, and the proposed method on two different datasets. We directly cite the results reported in \cite{Djelouah_2019_ICCV} and \cite{Rippel_2019_ICCV}. Please note that Djelouah\_ICCV2019 \cite{Djelouah_2019_ICCV} is designed for random-access scenarios and uses the autoregressive entropy model proposed in \cite{minnen2018joint}, while our method targets low-latency scenarios and just uses the fully-factorized (\cite{balle2016end}) and hyperprior (\cite{balle2018variational}) entropy model. Rippel\_ICCV2019 \cite{Rippel_2019_ICCV} is optimized for MS-SSIM but ours is optimized for MSE, PSNR results were not reported in \cite{Rippel_2019_ICCV}.}
  \label{fig_ICCV Curve} 
\end{figure*}
In the paper, we compare with two learned video compression methods of the state-of-the-art, \ie Wu\_ECCV2018 \cite{wu2018video} and DVC \cite{lu2018dvc}. Here, we also compare with other two latest learned methods, i.e. Djelouah\_ICCV2019 \cite{Djelouah_2019_ICCV} designed for random-access scenarios and Rippel\_ICCV2019 \cite{Rippel_2019_ICCV} targeting low-latency scenarios. From Fig.~\ref{fig_ICCV Curve} (b), we can observe that Djelouah\_ICCV2019 \cite{Djelouah_2019_ICCV} achieves better performance of $0.25\sim0.7$dB gain than our method in terms of PSNR on the MCL\_JCV dataset \cite{wang2016mcl}. Note that, their method is designed for random-access scenarios and integrates the autoregressive prior, proposed in \cite{minnen2018joint}, to predict the probabilities of quantized representations in entropy model. This autoregressive model has an obvious disadvantage of high decoding complexity even in parallel devices like GPU/TPU. From Fig.~\ref{fig_ICCV Curve} (c), we can observe that Rippel\_ICCV2019 \cite{Rippel_2019_ICCV} outperforms our method by about 0.005 in terms of MS-SSIM on the Xiph 1080p video dataset \cite{xiphdata}. Note that, their method is optimized directly for MS-SSIM, but ours is optimized for MSE. It requires our future work to optimize our model for MS-SSIM to achieve a better performance in MS-SSIM.
\subsection{Comparison with H.264 and H.265 in Other Settings}
\begin{figure*}
  \centering
  \subfigure[]{
    \includegraphics[width=2.2in]{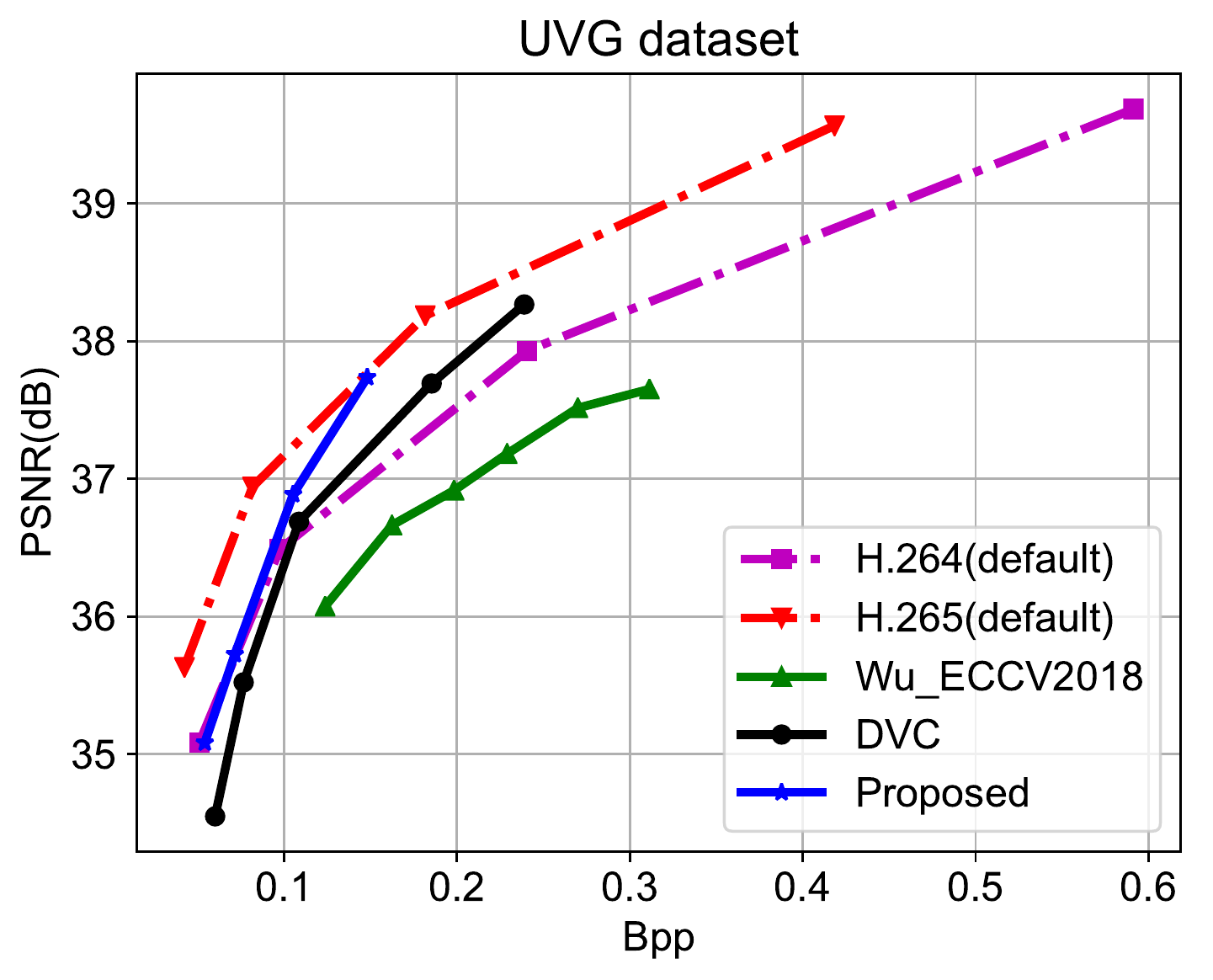}}
  \subfigure[]{
    \includegraphics[width=2.2in]{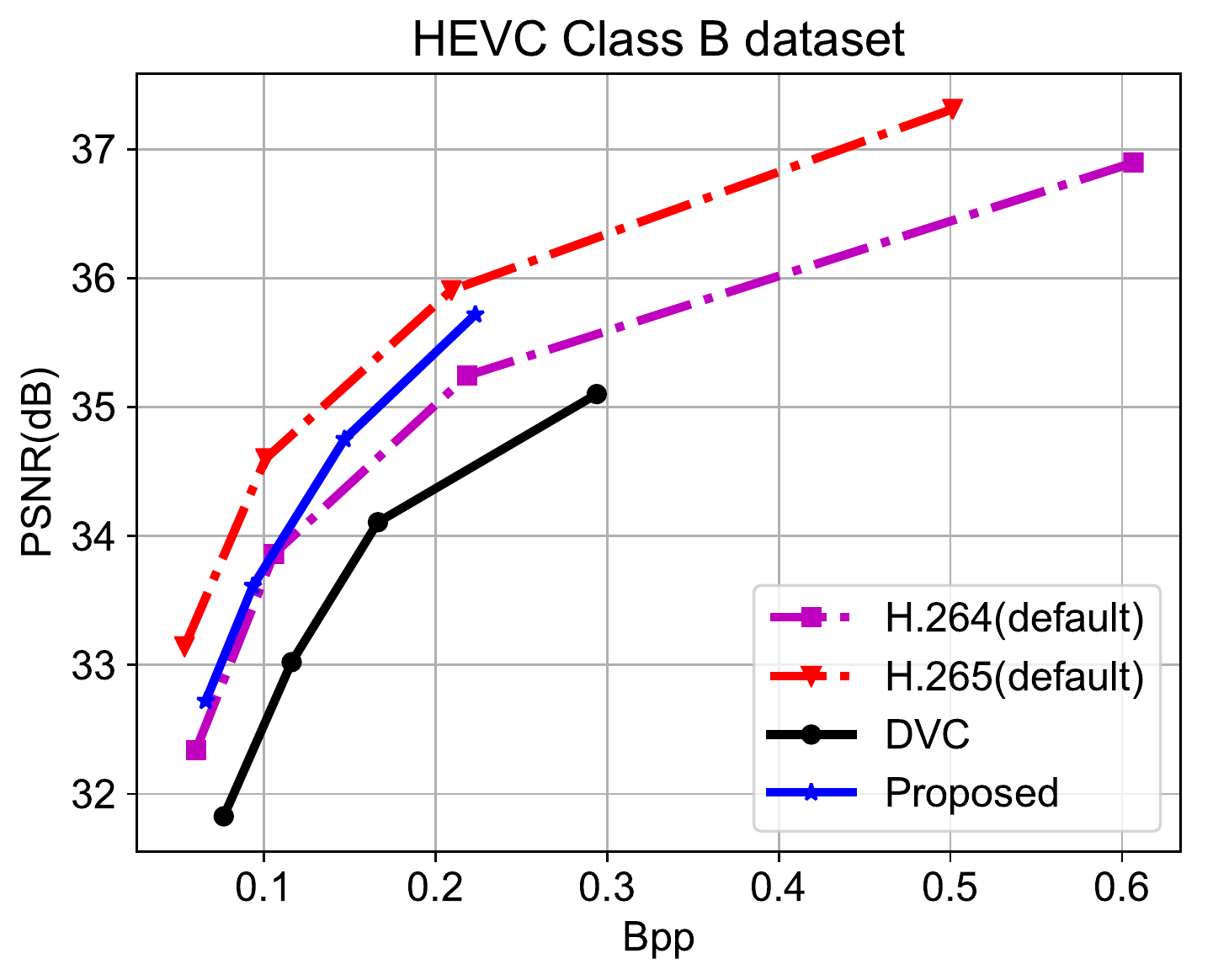}}
  \subfigure[]{
    \includegraphics[width=2.2in]{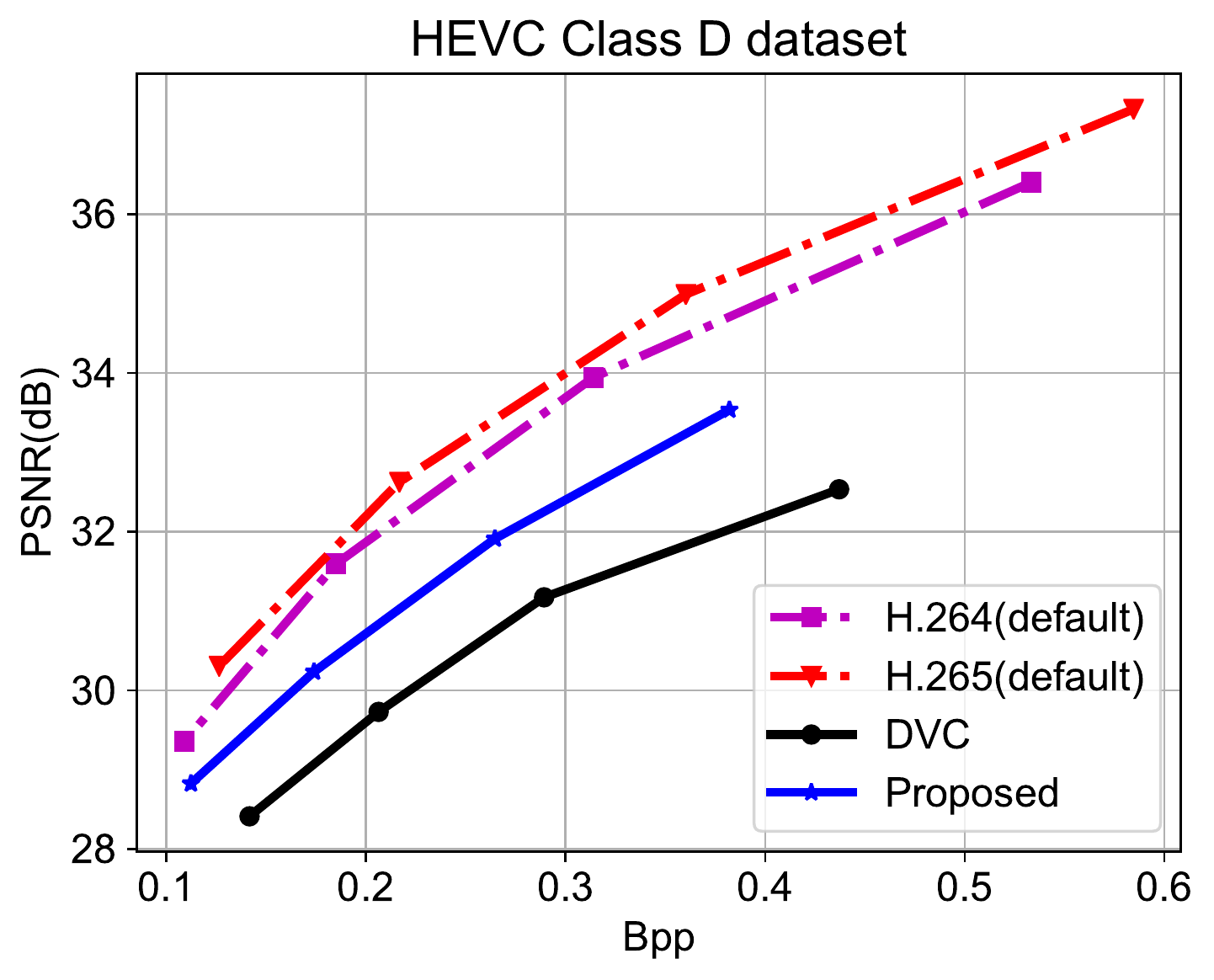}}
  \subfigure[]{
    \includegraphics[width=2.2in]{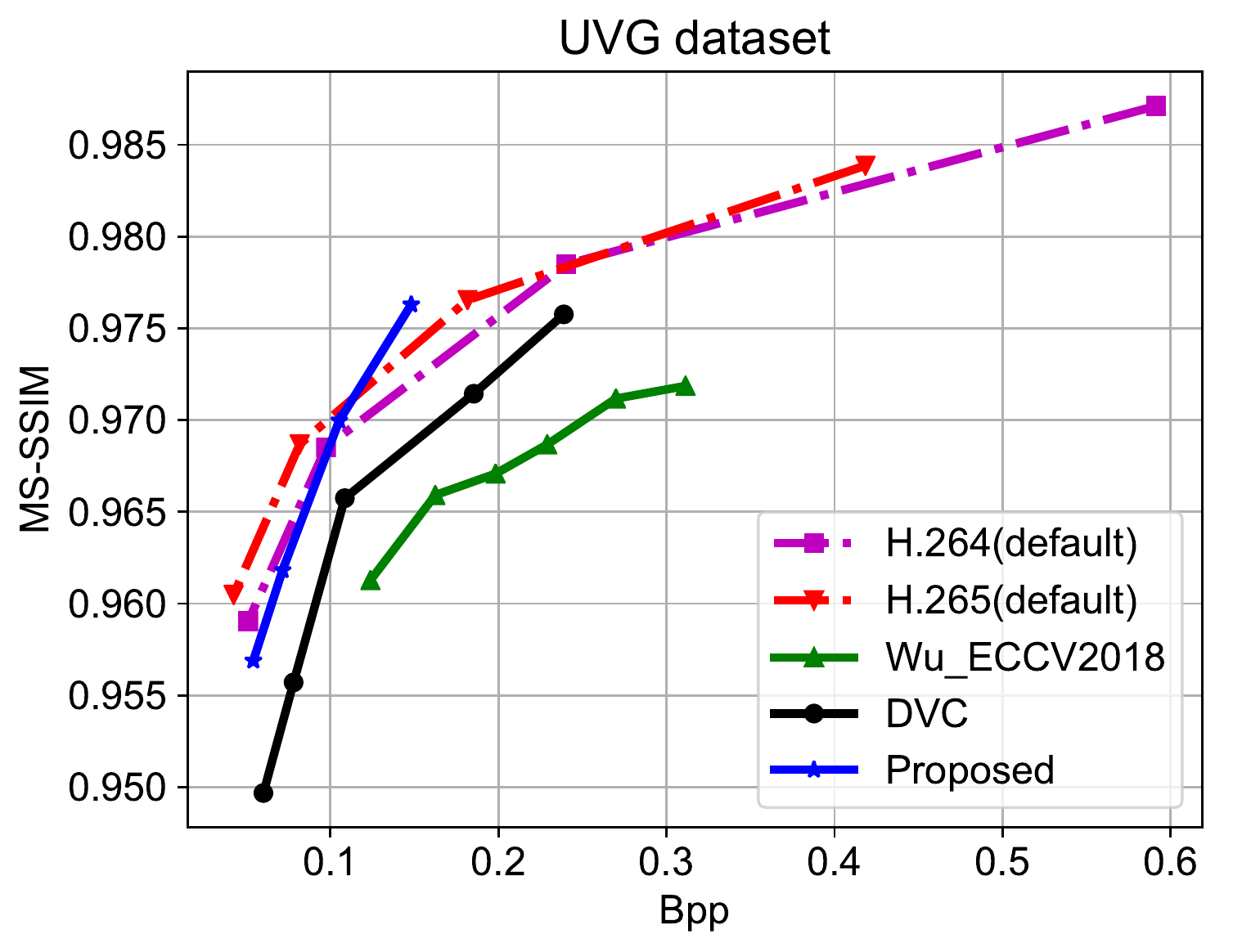}}
  \subfigure[]{
    \includegraphics[width=2.2in]{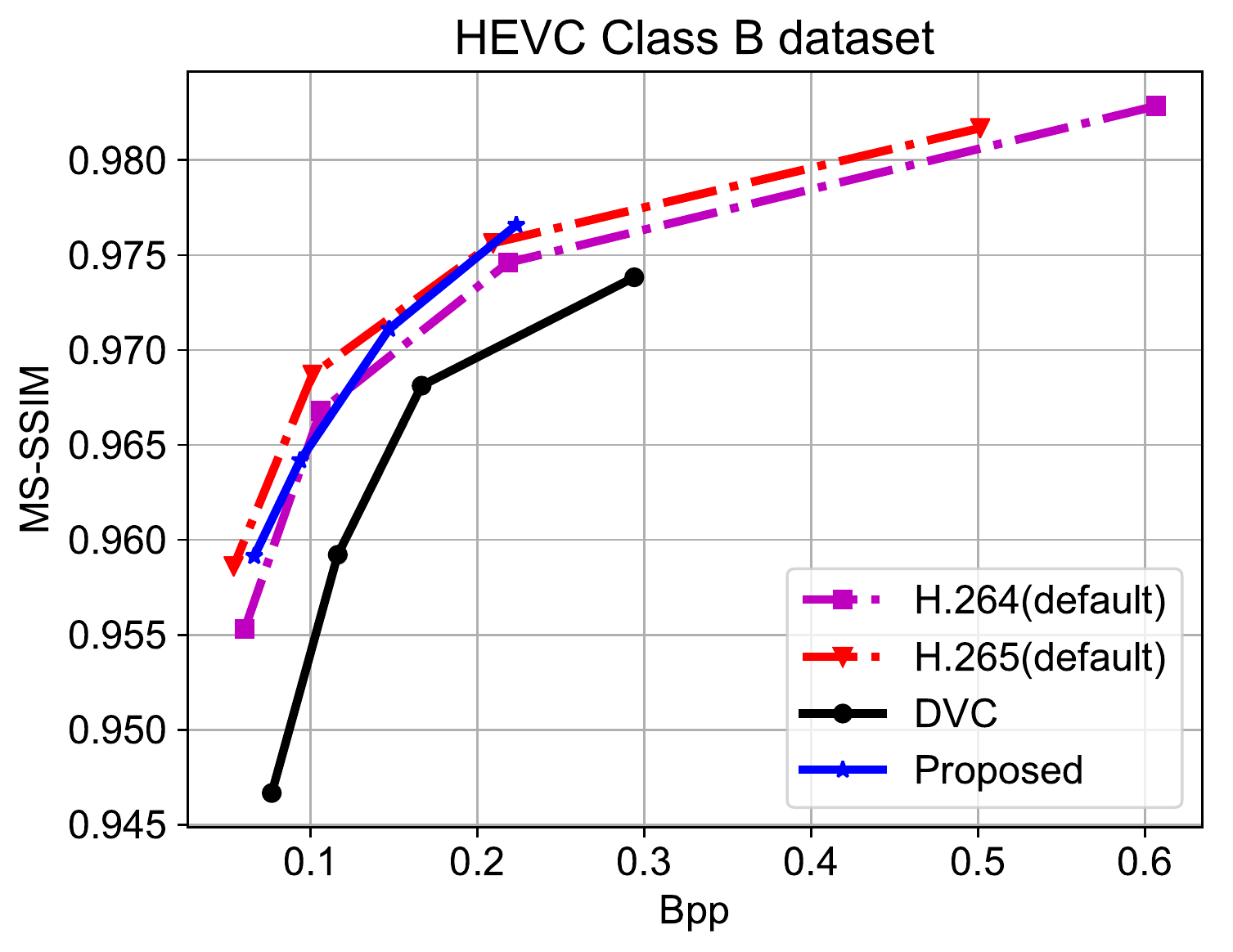}}
  \subfigure[]{
    \includegraphics[width=2.2in]{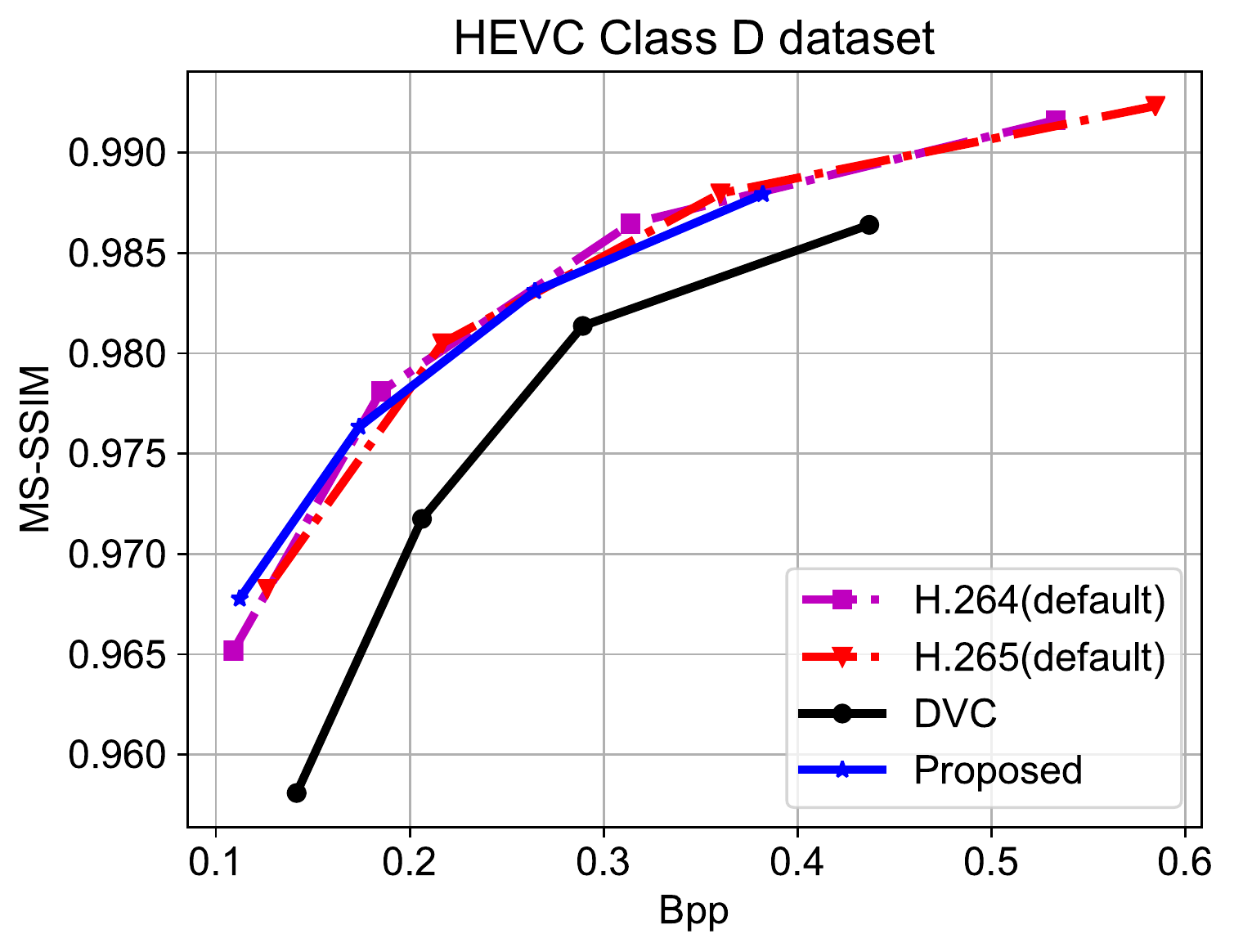}}
  \caption{Compression results on the three datasets using H.264, H.265, DVC \cite{lu2018dvc}, Wu's method \cite{wu2018video} and the proposed method. The settings of H.264 and H.265 are specified in the text. Top row: PSNR. Bottom row: MS-SSIM.}
  \label{fig_RD Curve} 
\vspace{-0.3cm}
\end{figure*}
In the paper, we compare with the results of H.264 and H.265 where the results are directly cited from \cite{lu2018dvc}. Note that the results are obtained by using the \texttt{veryfast} mode of x264 and x265 codecs, respectively. Here, we also compare with the results of H.264 and H.265 using other settings. Specifically, we use the following command lines for compressing a sequence \emph{Video.yuv} whose resolution is \emph{W}$\times$\emph{H} using x264 and x265 codecs,\\
\textit{ffmpeg -y -pix\_fmt yuv420p -s WxH -r FR -i Video.yuv -vframes N -c:v libx264 -crf Q -loglevel debug output.mkv}\\
\textit{ffmpeg -y -pix\_fmt yuv420p -s WxH -r FR -i Video.yuv -vframes N -c:v libx265 -x265-params ``crf=Q'' output.mkv}\\
where \emph{FR}, \emph{N}, \emph{Q} stand for the frame rate, the number of frames to be encoded, and the quality level, respectively.

Fig.\ \ref{fig_RD Curve} presents the compression results on the UVG dataset and the HEVC Class B and Class D datasets. It can be observed that our proposed method achieves competitive results than x264 in PSNR, and is on par with x265 in MS-SSIM.

\end{document}